\documentclass[a4paper,12pt]{article}

\usepackage[english]{babel}
\usepackage[latin1]{inputenc}

\usepackage{indentfirst}
\usepackage[dvips,final]{graphicx}
\usepackage{amsfonts,amsmath,amssymb,amsthm}

\usepackage{color}
\definecolor{oneblue}{rgb}{0,0.0,0.75}
\usepackage[colorlinks,
            urlcolor=oneblue,
            linkcolor=oneblue,
            citecolor=oneblue,
            bookmarksopen=false,
            pagebackref]{hyperref}

\hypersetup{
    pdftitle = Water waves generated by a moving bottom,
    pdfauthor = Denys Dutykh and Frédéric Dias,
    pdfsubject = Linear water waves
    pdfkeywords = {linear water waves, Cauchy-Poisson problem, tsunami, waves generation}
}

\usepackage[dvips,final]{graphicx}
\usepackage{xspace} 

\usepackage{fancyhdr}
\def\Dv{\mathbf{D}}
\def\kv{\mathbf{k}}
\def\uv{\mathbf{u}}
\def\qv{\mathbf{q}}
\def\nv{\mathbf{n}}
\def\R{\mathbb{R}}

\newcommand{\pd}[2]{\frac{\partial#1}{\partial#2}}
\newcommand{\od}[2]{\frac{d#1}{d#2}}
\newcommand{\tens}[1]{\mathrm{#1}}

\newcommand{\abs}[1]{\left|#1\right|}

\title{Water waves generated by a moving bottom}
\author{\href{http://www.cmla.ens-cachan.fr/\~dutykh}{Denys Dutykh%
\footnote{Centre de Mathématiques et de Leurs Applications, 
\'{E}cole Normale Supérieure de Cachan,
61, avenue du Président Wilson, 
94235 Cachan cedex, France}} \and
\href{http://www.cmla.ens-cachan.fr/\~dias}{Frédéric Dias\footnotemark[1]}}
\date{}

\begin{document}

\maketitle
\begin{abstract}
Tsunamis are often generated by a moving sea bottom. This paper deals with the case 
where the tsunami source is an earthquake. The linearized water-wave equations are 
solved analytically for various sea bottom motions. Numerical results based on 
the analytical solutions are shown for the free-surface profiles, the horizontal 
and vertical velocities as well as the bottom pressure. 
\end{abstract}
\tableofcontents

\section{Introduction}
Waves at the surface of a liquid can be generated by various mechanisms: wind blowing on the free surface,
wavemaker, moving disturbance on the bottom or the surface, or even inside the liquid, fall of an object 
into the liquid, liquid inside a moving container, etc. In this paper, we concentrate on
the case where the waves are created by a given motion of the bottom. One example is the generation of
tsunamis by a sudden seafloor deformation.

There are different natural phenomena that can lead to a tsunami. For example, one can mention submarine 
slumps, slides, volcanic explosions, etc. In this article we use a submarine faulting generation
mechanism as tsunami source. The resulting waves have some well-known features. For example, characteristic wavelengths 
are large and wave amplitudes are small compared with water depth. 

Two factors are usually necessary for an accurate modelling of tsunamis: information on the magnitude and distribution 
of the displacements caused by the earthquake, and a model of surface gravity waves generation resulting 
from this motion of the seafloor. Most studies of tsunami generation assume that the initial free-surface deformation 
is equal to the vertical displacement of the ocean bottom. The details of wave motion
are neglected during the time that the source operates. While this is often justified because the earthquake rupture
occurs very rapidly, there are some specific cases where the time scale of the bottom deformation may become an 
important factor. This was emphasized for example
by Trifunac and Todorovska \cite{todo}, who considered the generation of tsunamis by a slowly spreading uplift
of the seafloor and were able to explain some observations. During the 26 December 2004 Sumatra-Andaman event,
there was in the northern extent of the source a relatively slow faulting motion that led to significant vertical
bottom motion but left little record in the seismic data. It is interesting to point out that it is the inversion of
tide-gauge data from Paradip, the northernmost of the Indian east-coast stations, that led Neetu et al. \cite{indiens2}
to conclude that the source length was greater by roughly $30\%$ than the initial estimate of Lay et al. \cite{Lay}. 
Incidentally, the generation time is also longer for landslide tsunamis.

Our study is restricted to the water region where the incompressible Euler equations for potential flow can be 
linearized. The wave propagation away from the source can be investigated by shallow water models which may or may not
take into account nonlinear effects and frequency dispersion. Such models include the
Korteweg-de Vries equation \cite{KdV} for unidirectional propagation, nonlinear shallow-water equations
and Boussinesq-type models \cite{bouss,peregr,bona}.

Several authors have modeled the incompressible fluid layer as a special case of an elastic medium 
\cite{podyapolsk1, kajiura, gusyakov, aleksgus, gusyakov3}. In our opinion it 
may be convenient to model the liquid by an elastic material from a mathematical point of view, but it is questionable 
from a physical point of view. The crust was modeled as an
elastic isotropic half-space. This assumption will also be adopted in the present study.

The problem of tsunami generation has been considered by a number of authors: see for example \cite{carrier, driess, bradd}.  
The models discussed in these papers lack flexibility in terms of modelling the source due to the earthquake. The present paper
provides some extensions. A good review on the subject is \cite{sabatier}.

Here we essentially follow the framework proposed by Hammack \cite{Hammack} and others. The tsunami generation 
problem is reduced to a Cauchy-Poisson boundary value problem in a region of
constant depth. The main extensions given in the present paper consist in three-dimensional modelling and more realistic source 
models. This approach was followed recently in \cite{todo, todo2}, where the mathematical model was the same as in
\cite{Hammack} but the source was different.

Most analytical studies of linearized wave motion use integral transform methods. The complexity 
of the integral solutions forced many authors \cite{kajiura, keller} to use
asymptotic methods such as the method of stationary phase to estimate the far-field behaviour of the solutions. In the present 
study we have also obtained asymptotic formulas for integral solutions. They are useful from a qualitative point of view, 
but in practice it is better to use numerical integration formulas \cite{filon} that take into account the oscillatory nature
of the integrals. All the numerical results presented in this paper were obtained in this manner.

One should use asymptotic solutions with caution since they approximate exact solutions of the linearized problem. The relative 
importance of linear and nonlinear effects can be measured by the Stokes (or Ursell) number \cite{ursell}:
$$
  U := \frac{a/h}{(kh)^2} = \frac{a}{k^2h^3},
$$
where $k$ is a wave number, $a$ a typical wave amplitude and $h$ the water depth. For $U \gg 1$, the nonlinear effects 
control wave propagation and only nonlinear models are applicable. Ursell \cite{ursell}
proved that near the wave front $U$ behaves like
$$
  U \sim t^{\frac13}.
$$
Hence, regardless of how small nonlinear effects are initially, they will become important. 

Section 2 provides a description of the tsunami source when the source is an earthquake. In Section 3, we review
the water-wave equations and 
provide the analytical solution to the linearized problem in the fluid domain. Section 4 is devoted to
numerical results based on the analytical solution. 

\section{Source model}

The inversion of seismic wave data allows the reconstruction of
permanent deformations of the sea bottom following earthquakes. In
spite of the complexity of the seismic source and of the internal
structure of the earth, scientists have been relatively successful
in using simple models for the source. One of these models
is Okada's model \cite{Okada85}. Its description follows.

The fracture zones, along which the foci of earthquakes are to be
found, have been described in various papers. For example, it has
been suggested that Volterra's theory of dislocations might be the
proper tool for a quantitative description of these fracture zones
\cite{stek2}. This suggestion was made for the following reason. If
the mechanism involved in earthquakes and the fracture zones is
indeed one of fracture, discontinuities in the displacement
components across the fractured surface will exist. As dislocation
theory may be described as that part of the theory of elasticity
dealing with surfaces across which the displacement field is
discontinuous, the suggestion makes sense.

As is often done in mathematical physics, it is necessary for
simplicity's sake to make some assumptions. Here we neglect the
curvature of the earth, its gravity, temperature, magnetism,
non-homogeneity, and consider a semi-infinite medium, which is
homogeneous and isotropic. We further assume that the laws of
classical linear elasticity theory hold.

Several studies showed that the effect of earth curvature is
negligible for shallow events at distances of less than $20^\circ$
\cite{ben1,ben2,smylie}. The sensitivity to earth topography,
homogeneity, isotropy and half-space assumptions was studied and
discussed recently \cite{tim}. A commercially
available code, ABACUS, which is based on a finite element model
(FEM), was used. Six FEMs were constructed to test the sensitivity of
deformation predictions to each assumption. The author came to the conclusion
that the vertical layering of lateral inhomogeneity can sometimes
cause considerable effects on the deformation fields.

The usual boundary conditions for dealing with earth problems
require that the surface of the elastic medium (the earth) shall
be free from forces. The resulting mixed boundary-value problem was
solved a century ago \cite{volt}. Later, Steketee proposed an
alternative method to solve this problem using Green's functions
\cite{stek2}.

\subsection{Volterra's theory of dislocations}

In order to introduce the concept of dislocation and for
simplicity's sake, this section is devoted to the case of an entire
elastic space, as was done in the original paper by
Volterra \cite{volt}.

Let $O$ be the origin of a Cartesian coordinate system in an
infinite elastic medium, $x_i$ the Cartesian coordinates
$(i=1,2,3)$, and $\mathbf{e}_i$ a unit vector in the positive
$x_i-$direction. A force $\mathbf{F}=F \mathbf{e}_k$ at $O$
generates a displacement field $u_i^k(P,O)$ at point $P$, which is
determined by the well-known Somigliana tensor
\begin{equation}\label{somigliana}
  u_i^k(P,O) = \frac{F}{8\pi\mu} (\delta_{ik} r_{,\: nn}
  - \alpha r_{,\:ik}), \quad \mbox{with} \;\; \alpha=\frac{\lambda+\mu}{\lambda+2\mu}.
\end{equation}
In this relation $\delta_{ik}$ is the Kronecker delta, $\lambda$ and
$\mu$ are Lam\'e's constants, and $r$ is the distance from $P$ to
$O$. The coefficient $\alpha$ can be rewritten as
$\alpha=1/2(1-\nu)$, where $\nu$ is Poisson's ratio. Later we will
also use Young's modulus $E$, which is defined as
\[ E = \frac{\mu\,(3\lambda+2\mu)}{\lambda+\mu}. \]
The notation $r_{,\:i}$ means $\partial r/\partial x_i$ and the
summation convention applies.

The stresses due to the displacement field (\ref{somigliana}) are
easily computed from Hooke's law:

\begin{equation}\label{hook}
  \sigma_{ij} = \lambda\delta_{ij} u_{k,k} + \mu
  (u_{i,j}+u_{j,i}).
\end{equation}
One finds
\[
  \sigma_{ij}^k (P,O) = -\frac{\alpha F}{4\pi} \left(
  \frac{3 x_i x_j x_k}{r^5} + \frac{\mu}{\lambda+\mu}
  \frac{\delta_{ki}x_j + \delta_{kj}x_i - \delta_{ij}x_k}{r^3}
  \right).
\]
The components of the force per unit area on a surface element are
denoted as follows:
\[
  T_i^{k} = \sigma_{ij}^k \nu_j,
\]
where the $\nu_j$'s are the components of the normal to the
surface element. A Volterra dislocation is defined as a
surface $\Sigma$ in the elastic medium across which there is a
discontinuity $\Delta u_i$ in the displacement fields of the type
\begin{eqnarray}\label{displdef}
  \Delta u_i & = & u_i^+ - u_i^- = U_i + \Omega_{ij}x_j, \\
  \Omega_{ij}& = & -\Omega_{ji}.
\end{eqnarray}
Equation (\ref{displdef}) in which $U_i$ and $\Omega_{ij}$ are
constants is the well-known Weingarten relation which states that
the discontinuity $\Delta u_i$ should be of the type of a rigid body
displacement, thereby maintaining continuity of the components of
stress and strain across $\Sigma$.

The displacement field in an infinite elastic medium due to the
dislocation is then determined by Volterra's formula \cite{volt}
\begin{equation}\label{volt2}
  u_k(Q) = \frac{1}{F}\int\!\!\!\!\int
  \limits_{\!\!\!\!\!\!\!\Sigma} \Delta u_i
  T_i^{k} \, dS.
\end{equation}

Once the surface $\Sigma$ is given, the dislocation is essentially
determined by the six constants $U_i$ and $\Omega_{ij}$. Therefore
we also write
\begin{equation}\label{eldisloc}
  u_k(Q) = \frac{U_i}{F}\int\!\!\!\!\int
  \limits_{\!\!\!\!\!\!\!\Sigma} \sigma_{ij}^k (P,Q) \nu_j dS +
  \frac{\Omega_{ij}}{F}\int\!\!\!\!\int
  \limits_{\!\!\!\!\!\!\!\Sigma} \{x_j\sigma_{il}^k(P,Q) - x_i
  \sigma_{jl}^k(P,Q)\}\nu_l dS,
\end{equation}
where $\Omega_{ij}$ takes only the values $\Omega_{12}$,
$\Omega_{23}$, $\Omega_{31}$. Following Volterra \cite{volt} and
Love \cite{love} we call each of the six integrals in
(\ref{eldisloc}) an elementary dislocation.

It is clear from (\ref{volt2}) and (\ref{eldisloc}) that the
computation of the displacement field $u_k(Q)$ is performed as
follows. A force $F\mathbf{e}_k$ is applied at $Q$, and the stresses
$\sigma_{ij}^k(P,Q)$ that this force generates are computed at the
points $P(x_i)$ on $\Sigma$. In particular the components of the
force on $\Sigma$ are computed. After multiplication with prescribed
weights of magnitude $\Delta u_i$ these forces are integrated over
$\Sigma$ to give the displacement component in $Q$ due to the
dislocation on $\Sigma$.

\subsection{Dislocations in elastic half-space}

When the case of an elastic half-space is considered, equation
(\ref{volt2}) remains valid, but we have to replace $\sigma_{ij}^k$ in $T_i^k$ by
another tensor $\omega_{ij}^k$. This can be explained by the fact
that the elementary solutions for a half-space are different from
Somigliana solution (\ref{somigliana}).

The $\omega_{ij}^k$ can be obtained from the displacements
corresponding to nuclei of strain in a half-space through relation
(\ref{hook}). Steketee showed a method of obtaining the six
$\omega_{ij}^k$ fields by using a Green's function and derived
$\omega_{12}^k$, which is relevant to a vertical strike-slip fault (see below).
Maruyama derived the remaining five functions \cite{maru}.

It is interesting to mention here that historically these solutions
were first derived in a straightforward manner by Mindlin
\cite{mindl1,mindl2}, who gave explicit expressions of the
displacement and stress fields for half-space nuclei of strain
consisting of single forces with and without moment. It is only
necessary to write the single force results since the other forms
can be obtained by taking appropriate derivatives. The method
consists in finding the displacement field in Westergaard's form of
the Galerkin vector \cite{wester}. This vector is then determined by
taking a linear combination of some biharmonic elementary solutions.
The coefficients are chosen to satisfy boundary and equilibrium
conditions. These solutions were also derived by Press in a slightly
different manner \cite{press}.

\begin{figure}[htbp]
\begin{center}
\unitlength 1mm
\begin{picture}(71.25,68.75)(0,0)

\linethickness{0.50mm}
\put(9.38,0.00){\line(0,1){68.75}}
\put(9.38,68.75){\vector(0,1){0.12}}

\put(4.38,68.13){\makebox(0,0)[cc]{$x_3$}}

\linethickness{0.50mm}
\multiput(9.38,45.63)(0.37,-0.12){151}{\line(1,0){0.37}}
\put(65.00,27.50){\vector(3,-1){0.12}}

\linethickness{0.15mm}
\multiput(9.38,15.00)(0.37,-0.12){109}{\line(1,0){0.37}}

\put(64.38,25.63){\makebox(0,0)[cc]{$x_1$}}

\linethickness{0.50mm}
\multiput(9.38,45.63)(0.22,0.12){120}{\line(1,0){0.22}}
\put(35.63,60.00){\vector(2,1){0.12}}

\put(35.63,63.13){\makebox(0,0)[cc]{$x_2$}}

\put(4.38,45.63){\makebox(0,0)[cc]{$O$}}

\put(0.63,15.00){\makebox(0,0)[cc]{$x_3=-d$}}

\linethickness{0.15mm}
\multiput(49.38,1.87)(0.21,0.12){73}{\line(1,0){0.21}}

\linethickness{0.15mm}
\put(65.00,10.63){\line(0,1){8.13}}

\linethickness{0.15mm}
\multiput(49.38,1.88)(0.12,0.13){130}{\line(0,1){0.13}}

\linethickness{0.15mm}
\multiput(25.00,31.88)(0.37,-0.12){109}{\line(1,0){0.37}}

\linethickness{0.15mm}
\multiput(9.38,15.00)(0.12,0.13){130}{\line(0,1){0.13}}

\put(60.00,10.00){\makebox(0,0)[cc]{$\delta$}}

\linethickness{0.15mm}
\multiput(56.74,7.12)(0.12,-0.99){1}{\line(0,-1){0.99}}
\multiput(56.50,8.07)(0.12,-0.48){2}{\line(0,-1){0.48}}
\multiput(56.15,8.96)(0.12,-0.30){3}{\line(0,-1){0.30}}

\put(27.50,5.63){\makebox(0,0)[cc]{$L$}}

\put(53.13,11.88){\makebox(0,0)[cc]{$W$}}

\linethickness{0.45mm}
\multiput(32.50,16.88)(0.34,-0.12){26}{\line(1,0){0.34}}
\put(41.25,13.75){\vector(3,-1){0.12}}

\linethickness{0.45mm}
\multiput(32.50,16.88)(0.12,0.15){42}{\line(0,1){0.15}}
\put(37.50,23.13){\vector(3,4){0.12}}

\put(40.00,11.25){\makebox(0,0)[cc]{$U_1$}}

\put(40.00,20.63){\makebox(0,0)[cc]{$U_2$}}

\linethickness{0.45mm}
\multiput(26.88,23.13)(0.12,-0.13){47}{\line(0,-1){0.13}}
\put(26.88,23.13){\vector(-1,1){0.12}}

\put(22.50,20.63){\makebox(0,0)[cc]{$U_3$}}

\linethickness{0.20mm}
\multiput(28.75,56.25)(0.37,-0.12){115}{\line(1,0){0.37}}

\linethickness{0.20mm}
\multiput(52.50,31.25)(0.20,0.12){94}{\line(1,0){0.20}}

\put(49.38,40.00){\makebox(0,0)[cc]{Free surface}}

\end{picture}
\end{center}
  \caption{Coordinate system adopted in this study and
geometry of the source model}\label{fig:okad}
\end{figure}
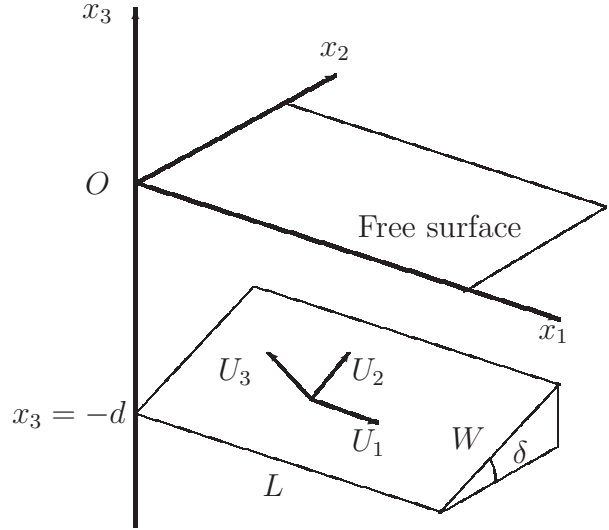

Here, we take the Cartesian coordinate system shown in
Figure~\ref{fig:okad}. The elastic medium occupies the region $x_3\leq
0$ and the $x_1-$axis is taken to be parallel to the strike direction
of the fault. In this coordinate system,
$u_i^j(x_1,x_2,x_3;\xi_1,\xi_2,\xi_3)$ is the $i$th component of the
displacement at $(x_1,x_2,x_3)$ due to the $j$th direction point
force of magnitude $F$ at $(\xi_1,\xi_2,\xi_3)$. It can be expressed
as follows \cite{Okada85,mindl1,press,okada92}:
\begin{eqnarray}\label{okada1}
  u_i^j (x_1,x_2,x_3) & = & u_{iA}^j(x_1,x_2,-x_3) -
  u_{iA}^j(x_1,x_2,x_3) \\
  & & + u_{iB}^j(x_1,x_2,x_3) + x_3
  u_{iC}^j(x_1,x_2,x_3), \nonumber
\end{eqnarray}
where
\begin{eqnarray*}
  u_{iA}^j & = & \frac{F}{8\pi\mu}\left((2-\alpha)\frac{\delta_{ij}}{R}
  + \alpha \frac{R_iR_j}{R^3}\right), \\
  u_{iB}^j & = & \frac{F}{4\pi\mu}\Biggl(
  \frac{\delta_{ij}}{R} + \frac{R_iR_j}{R^3} +
  \frac{1-\alpha}{\alpha}\Bigl[
  \frac{\delta_{ij}}{R+R_3} +\\ & & + \frac{R_i\delta_{j3}-
  R_j\delta_{i3}(1-\delta_{j3})}{R(R+R_3)} -
  \frac{R_iR_j}{R(R+R_3)^2}(1-\delta_{i3})(1-\delta_{j3})
  \Bigr]\Biggr), \\
    u_{iC}^j & = & \frac{F}{4\pi\mu}(1-2\delta_{i3})
    \left(
    (2-\alpha)\frac{R_i\delta_{j3}-R_j\delta_{i3}}{R^3} +
    \alpha\xi_3\left[
    \frac{\delta_{ij}}{R^3} - 3\frac{R_iR_j}{R^5}
    \right]\right).
\end{eqnarray*}
In these expressions $R_1 = x_1-\xi_1$, $R_2=x_2-\xi_2$,
$R_3=-x_3-\xi_3$ and $R^2 = R_1^2 + R_2^2 + R_3^2$.

The first term in equation (\ref{okada1}), $u_{iA}^j(x_1,x_2,-x_3)$,
is the well-known Somigliana tensor, which represents the
displacement field due to a single force placed at
$(\xi_1,\xi_2,\xi_3)$ in an infinite medium \cite{love}. The second
term also looks like a Somigliana tensor. This term corresponds to a
contribution from an image source of the given point force placed at
$(\xi_1,\xi_2,-\xi_3)$ in the infinite medium. The third term,
$u_{iB}^j(x_1,x_2,x_3)$, and $u_{iC}^j(x_1,x_2,x_3)$ in the fourth
term are naturally depth dependent. When $x_3$ is set equal to zero
in equation (\ref{okada1}), the first and the second terms cancel
each other, and the fourth term vanishes. The remaining term,
$u_{iB}^j(x_1,x_2,0)$, reduces to the formula for the surface
displacement field due to a point force in a half-space
\cite{Okada85}:
\[
\left\{%
\begin{array}{ll}
    u_1^1 = \frac{F}{4\pi\mu}\left(
    \frac1R + \frac{(x_1-\xi_1)^2}{R^3} + \frac{\mu}{\lambda+\mu}
    \left[
    \frac{1}{R-\xi_3} - \frac{(x_1-\xi_1)^2}{R(R-\xi_3)^2}\right]\right), & \\
    u_2^1 = \frac{F}{4\pi\mu}(x_1-\xi_1)(x_2-\xi_2)\left(
    \frac1{R^3} - \frac{\mu}{\lambda+\mu}\frac1{R(R-\xi_3)^2}
    \right), &  \\
    u_3^1 = \frac{F}{4\pi\mu}(x_1-\xi_1)\left(
    -\frac{\xi_3}{R^3} - \frac{\mu}{\lambda+\mu} \frac1{R(R-\xi_3)}
    \right), &
\end{array}%
\right.
\]

\[
\left\{%
\begin{array}{ll}
    u_1^2 = \frac{F}{4\pi\mu}(x_1-\xi_1)(x_2-\xi_2)\left(
    \frac1{R^3} - \frac{\mu}{\lambda+\mu}\frac{1}{R(R-\xi_3)^2}\right), & \\
    u_2^2 = \frac{F}{4\pi\mu}\left(
    \frac1R + \frac{(x_2-\xi_2)^2}{R^3} + \frac{\mu}{\lambda+\mu}
    \left[
    \frac{1}{R-\xi_3} - \frac{(x_2-\xi_2)^2}{R(R-\xi_3)^2}\right]\right), &  \\
    u_3^2 = \frac{F}{4\pi\mu}(x_2-\xi_2)\left(
    -\frac{\xi_3}{R^3} - \frac{\mu}{\lambda+\mu} \frac1{R(R-\xi_3)}
    \right), &
\end{array}%
\right.
\]

\[
\left\{%
\begin{array}{ll}
    u_1^3 = \frac{F}{4\pi\mu}(x_1-\xi_1)\left(
    -\frac{\xi_3}{R^3} + \frac{\mu}{\lambda+\mu}\frac{1}{R(R-\xi_3)}\right), & \\
    u_2^3 = \frac{F}{4\pi\mu}(x_2-\xi_2)\left(
    -\frac{\xi_3}{R^3} + \frac{\mu}{\lambda+\mu}\frac{1}{R(R-\xi_3)}\right), &  \\
    u_3^3 = \frac{F}{4\pi\mu}\left(
    \frac1{R} + \frac{\xi_3^2}{R^3} + \frac{\mu}{\lambda+\mu}\frac1R
    \right). &
\end{array}%
\right.
\]
In these formulas $R^2 = (x_1-\xi_1)^2 + (x_2-\xi_2)^2 + \xi_3^2$.

In order to obtain the displacements due to the dislocation we need
to calculate the corresponding $\xi_k$-derivatives of the point
force solution (\ref{okada1}) and to insert them in Volterra's
formula (\ref{volt2})
\[
  u_i = \frac1F\int\!\!\!\!\int
  \limits_{\!\!\!\!\!\!\!\Sigma} \Delta u_j
  \left[
  \lambda\delta_{jk} \pd{u_i^n}{\xi_n} +
  \mu\left(\pd{u_i^j}{\xi_k} + \pd{u_i^k}{\xi_j}
  \right)\right]\nu_k \,dS.
\]
The $\xi_k$-derivatives are expressed as follows:
\begin{eqnarray*}
  \pd{u_i^j}{\xi_k} (x_1,x_2,x_3) & = & \pd{u_{iA}^j}{\xi_k}(x_1,x_2,-x_3) -
  \pd{u_{iA}^j}{\xi_k}(x_1,x_2,x_3) +\\ & & + \pd{u_{iB}^j}{\xi_k}(x_1,x_2,x_3) + x_3
  \pd{u_{iC}^j}{\xi_k}(x_1,x_2,x_3),
\end{eqnarray*}
with
\begin{eqnarray*}
  \pd{u_{iA}^j}{\xi_k} & = & \frac{F}{8\pi\mu}\left(
  (2-\alpha)\frac{R_k}{R^3}\delta_{ij} - \alpha
  \frac{R_i\delta_{jk}+R_j\delta_{ik}}{R^3} + 3\alpha
  \frac{R_i R_j R_k}{R^5}
  \right), \\
  \pd{u_{iB}^j}{\xi_k} & = & \frac{F}{4\pi\mu}\left(
  -\frac{R_i\delta_{jk} + R_j\delta_{ik} - R_k\delta_{ij}}{R^3} +
  3\frac{R_iR_jR_k}{R^5} \right. +\\ & & + \frac{1-\alpha}{\alpha}\Bigl[
  \frac{\delta_{3k}R + R_k}{R(R+R_3)^2}\delta_{ij} -
  \frac{\delta_{ik}\delta_{j3} -
  \delta_{jk}\delta_{i3}(1-\delta_{j3})}{R(R+R_3)} +\\ & & +
  \bigl(R_i\delta_{j3} - R_j\delta_{i3}(1-\delta_{j3})\bigr)
  \frac{\delta_{3k}R^2+R_k(2R+R_3)}{R^3(R+R_3)^2} +\\ & & \left. +
  (1-\delta_{i3})(1-\delta_{j3})\bigl(
  \frac{R_i\delta_{jk}+R_j\delta_{ik}}{R(R+R_3)^2} -
  R_iR_j\frac{2\delta_{3k}R^2 + R_k(3R+R_3)}{R^3(R+R_3)^3}
  \bigr)\Bigr]\right), \\
  \pd{u_{iC}^j}{\xi_k} & = & \frac{F}{4\pi\mu}(1-2\delta_{i3})\biggl(
  (2-\alpha)\Bigl[
  \frac{\delta_{jk}\delta_{i3}-\delta_{ik}\delta_{j3}}{R^3} +
  \frac{3R_k(R_i\delta_{j3}-R_j\delta_{i3})}{R^5}\Bigr] + \\& & +
  \alpha\delta_{3k}\Bigl[\frac{\delta_{ij}}{R^3} -
  \frac{3R_iR_j}{R^5}\Bigr] + 3\alpha\xi_3\Bigl[
  \frac{R_i\delta_{jk}+R_j\delta_{ik}+R_k\delta_{ij}}{R^5} -
  \frac{5R_iR_jR_k}{R^7}\Bigr]\biggr).
\end{eqnarray*}

\subsection{Finite rectangular source}
Let us now consider a more practical problem. We define the
elementary dislocations $U_1$, $U_2$ and $U_3$, corresponding to
the strike-slip, dip-slip and tensile components of an arbitrary
dislocation. In Figure \ref{fig:okad} each vector represents the
direction of the elementary faults. The vector $\Dv$ is the
so-called Burger's vector, which shows how both sides of the fault
are spread out: $\Dv = \uv^+ - \uv^-$.

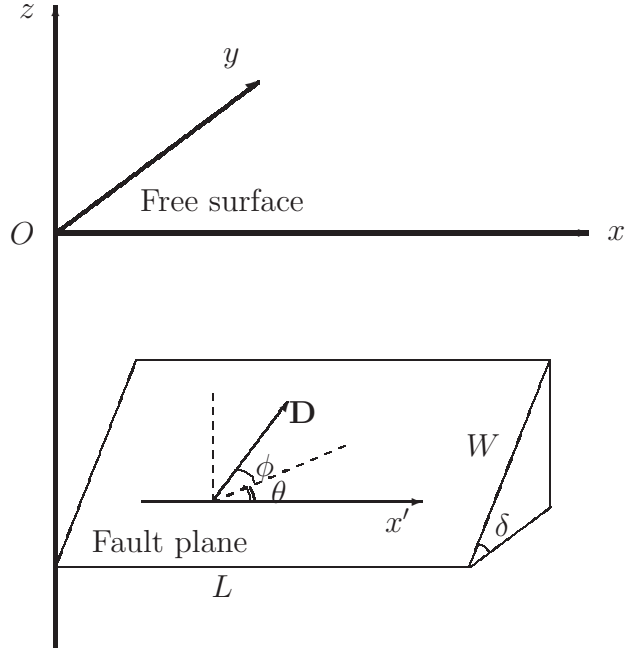
\begin{figure}[htbp]
\begin{center}
\unitlength 1mm
\def\Dv{\mathbf{D}}

\begin{picture}(83.75,84.38)(0,0)

\linethickness{0.55mm}
\put(10.00,54.38){\line(1,0){70.00}}
\put(80.00,54.38){\vector(1,0){0.12}}

\linethickness{0.55mm}
\multiput(10.00,54.38)(0.16,0.12){167}{\line(1,0){0.16}}
\put(36.25,74.38){\vector(4,3){0.12}}

\put(83.75,54.38){\makebox(0,0)[cc]{$x$}}

\put(33.13,77.50){\makebox(0,0)[cc]{$y$}}

\linethickness{0.45mm}
\put(10.00,-0.63){\line(0,1){85.00}}
\put(10.00,84.38){\vector(0,1){0.12}}

\put(6.25,83.75){\makebox(0,0)[cc]{$z$}}

\put(5.63,54.38){\makebox(0,0)[cc]{$O$}}

\linethickness{0.15mm}
\multiput(10.00,10.00)(0.12,0.31){89}{\line(0,1){0.31}}

\linethickness{0.15mm}
\put(10.00,10.00){\line(1,0){54.38}}

\linethickness{0.15mm}
\multiput(64.38,10.00)(0.12,0.31){89}{\line(0,1){0.31}}

\linethickness{0.15mm}
\multiput(64.38,10.00)(0.16,0.12){68}{\line(1,0){0.16}}

\linethickness{0.15mm}
\put(75.01,18.13){\line(0,1){19.37}}

\put(68.76,15.63){\makebox(0,0)[cc]{$\delta$}}

\linethickness{0.15mm}
\multiput(66.37,12.73)(0.12,-0.12){6}{\line(1,0){0.12}}
\multiput(65.42,13.10)(0.32,-0.12){3}{\line(1,0){0.32}}

\linethickness{0.25mm}
\multiput(30.63,18.75)(0.12,0.16){83}{\line(0,1){0.16}}
\put(40.63,31.88){\vector(3,4){0.12}}

\linethickness{0.15mm}
\multiput(30.63,18.75)(1.84,0.79){10}{\multiput(0,0)(0.31,0.13){3}{\line(1,0){0.31}}}

\put(25.01,13.13){\makebox(0,0)[cc]{Fault plane}}

\put(31.88,58.76){\makebox(0,0)[cc]{Free surface}}

\put(42.51,30.63){\makebox(0,0)[cc]{$\Dv$}}

\linethickness{0.15mm}
\put(21.26,18.75){\line(1,0){36.87}}
\put(58.13,18.75){\vector(1,0){0.12}}

\put(55.01,16.25){\makebox(0,0)[cc]{$x'$}}

\linethickness{0.15mm}
\multiput(35.93,21.84)(0.13,-0.25){2}{\line(0,-1){0.25}}
\multiput(35.47,22.29)(0.12,-0.11){4}{\line(1,0){0.12}}
\multiput(34.84,22.66)(0.21,-0.12){3}{\line(1,0){0.21}}
\multiput(34.08,22.93)(0.38,-0.14){2}{\line(1,0){0.38}}

\put(37.51,23.13){\makebox(0,0)[cc]{$\phi$}}

\linethickness{0.15mm}
\multiput(35.55,19.64)(0.08,-0.68){1}{\line(0,-1){0.68}}
\multiput(35.26,20.29)(0.15,-0.32){2}{\line(0,-1){0.32}}
\multiput(34.76,20.86)(0.12,-0.14){4}{\line(0,-1){0.14}}

\linethickness{0.15mm}
\multiput(36.00,19.81)(0.12,-0.57){2}{\line(0,-1){0.57}}
\multiput(35.53,20.78)(0.12,-0.24){4}{\line(0,-1){0.24}}

\put(39.38,20.00){\makebox(0,0)[cc]{$\theta$}}

\put(49.38,14.38){\makebox(0,0)[cc]{}}

\linethickness{0.15mm}
\multiput(30.63,18.75)(0,1.92){8}{\line(0,1){0.96}}

\put(31.88,7.50){\makebox(0,0)[cc]{$L$}}

\put(66.26,26.25){\makebox(0,0)[cc]{$W$}}

\linethickness{0.15mm}
\put(20.63,37.50){\line(1,0){54.38}}

\end{picture}
\end{center}
  \caption{Geometry of the source model and orientation
  of Burger's vector $\Dv$}\label{fig:2}
\end{figure}

A general dislocation can be determined by three angles: the dip
angle $\delta$ of the fault $(0\le\delta\le\pi)$, the slip or rake angle $\theta$ $(0\le\theta\le\pi)$, and the angle
$\phi$ between the fault plane and Burger's vector $\Dv$. When dealing with a
geophysical application, an additional angle, the azimuth or strike, is introduced in order to provide an orientation of the fault.
The general situation is schematically described in Figure \ref{fig:2}.

For a finite rectangular fault with length $L$ and width $W$
occurring at depth $d$ (Figure \ref{fig:2}), the deformation field
can be evaluated analytically by a change of variables and by
integrating over the rectangle. This was done by several
authors \cite{Okada85,okada92,chin,sat74,iwa79}. Here we give the
results of their computations. The final results are represented below in
compact form, using Chinnery's notation $\|$ to
represent the substitution
\[
  f(\xi,\eta)\| = f(x,p) - f(x,p-W) - f(x-L,p) + f(x-L,p-W), 
\]
where $p = y\cos\delta + d\sin\delta$. Next we introduce the notation
\[
  q = y\sin\delta - d\cos\delta, \quad \tilde{y} = \eta\cos\delta + q\sin\delta,
 \quad \tilde{d} = \eta\sin\delta - q\cos\delta
\]
and
\[
  R^2 = \xi^2 + \eta^2 + q^2 = \xi^2 + \tilde{y}^2 + \tilde{d}^2, \quad
  X^2 = \xi^2 + q^2.
\]

The quantities $U_1$, $U_2$ and $U_3$ are linked to Burger's vector
through the identities
\[
  U_1 = |\Dv| \cos\phi\cos\theta, \quad
  U_2 = |\Dv| \cos\phi\sin\theta, \quad
  U_3 = |\Dv| \sin\phi.
\]
For a strike-slip dislocation, one has
\begin{eqnarray*}
  u_1 & = & -\frac{U_1}{2\pi}\left.\left(
  \frac{\xi q}{R(R+\eta)} + \arctan\frac{\xi\eta}{qR} +
  I_1\sin\delta\right)\right\|, \\
  u_2 & = & -\frac{U_1}{2\pi}\left.\left(
  \frac{\tilde{y} q}{R(R+\eta)} + \frac{q\cos\delta}{R+\eta} +
  I_2\sin\delta\right)\right\|, \\
  u_3 & = & -\frac{U_1}{2\pi}\left.\left(
  \frac{\tilde{d} q}{R(R+\eta)} + \frac{q\sin\delta}{R+\eta} +
  I_4\sin\delta\right)\right\|.
\end{eqnarray*}
For a dip-slip dislocation, one has
\begin{eqnarray*}
  u_1 & = & -\frac{U_2}{2\pi}\left.\left(
  \frac{q}{R} - I_3\sin\delta\cos\delta
  \right)\right\|, \\
  u_2 & = & -\frac{U_2}{2\pi}\left.\left(
  \frac{\tilde{y} q}{R(R+\xi)} +
  \cos\delta\arctan\frac{\xi\eta}{qR} - I_1\sin\delta\cos\delta
  \right)\right\|, \\
  u_3 & = & -\frac{U_2}{2\pi}\left.\left(
  \frac{\tilde{d} q}{R(R+\xi)} +
  \sin\delta\arctan\frac{\xi\eta}{qR} - I_5\sin\delta\cos\delta
  \right)\right\|.
\end{eqnarray*}
For a tensile fault dislocation, one has
\begin{eqnarray*}
  u_1 & = & \frac{U_3}{2\pi}\left.\left(
  \frac{q^2}{R(R+\eta)} - I_3\sin^2\delta
  \right)\right\|, \\
  u_2 & = & \frac{U_3}{2\pi}\left.\left(
  \frac{-\tilde{d} q}{R(R+\xi)} - \sin\delta\left[
  \frac{\xi q}{R(R+\eta)} - \arctan\frac{\xi\eta}{qR}
  \right] - I_1\sin^2\delta
  \right)\right\|, \\
  u_3 & = & \frac{U_3}{2\pi}\left.\left(
  \frac{\tilde{y} q}{R(R+\xi)} + \cos\delta\left[
  \frac{\xi q}{R(R+\eta)} - \arctan\frac{\xi\eta}{qR}
  \right] - I_5\sin^2\delta
  \right)\right\|.
\end{eqnarray*}
The terms $I_1,\dots,I_5$ are given by
\begin{eqnarray*}
  I_1 & = &
  -\frac{\mu}{\lambda+\mu}\frac{\xi}{(R+\tilde{d})\cos\delta} -
  \tan\delta I_5, \\
  I_2 & = & -\frac{\mu}{\lambda+\mu}\log(R+\eta) - I_3, \\
  I_3 & = & \frac{\mu}{\lambda+\mu}\left[
  \frac1{\cos\delta}\frac{\tilde{y}}{R+\tilde{d}} - \log(R+\eta)
  \right] + \tan\delta I_4, \\
  I_4 & = & \frac{\mu}{\mu+\lambda}\frac1{\cos\delta}\left(
  \log(R+\tilde{d}) - \sin\delta \log(R+\eta)
  \right), \\
  I_5 & = & \frac{\mu}{\lambda+\mu}\frac2{\cos\delta}
  \arctan\frac{\eta(X+q\cos\delta)+X(R+X)\sin\delta}{\xi(R+X)\cos\delta},
\end{eqnarray*}
and if $\cos\delta=0$,
\begin{eqnarray*}
  I_1 & = & -\frac{\mu}{2(\lambda+\mu)}
  \frac{\xi q}{(R+\tilde{d})^2}, \\
  I_3 & = & \frac{\mu}{2(\lambda+\mu)} \left[
  \frac{\eta}{R+\tilde{d}} + \frac{\tilde{y} q}{(R+\tilde{d})^2} -
  \log(R+\eta)\right], \\
  I_4 & = & -\frac{\mu}{\lambda+\mu} \frac{q}{R+\tilde{d}}, \\
  I_5 & = & -\frac{\mu}{\lambda+\mu} \frac{\xi\sin\delta}{R+\tilde{d}}.
\end{eqnarray*}

\begin{figure}
  \includegraphics[width=0.9\linewidth]{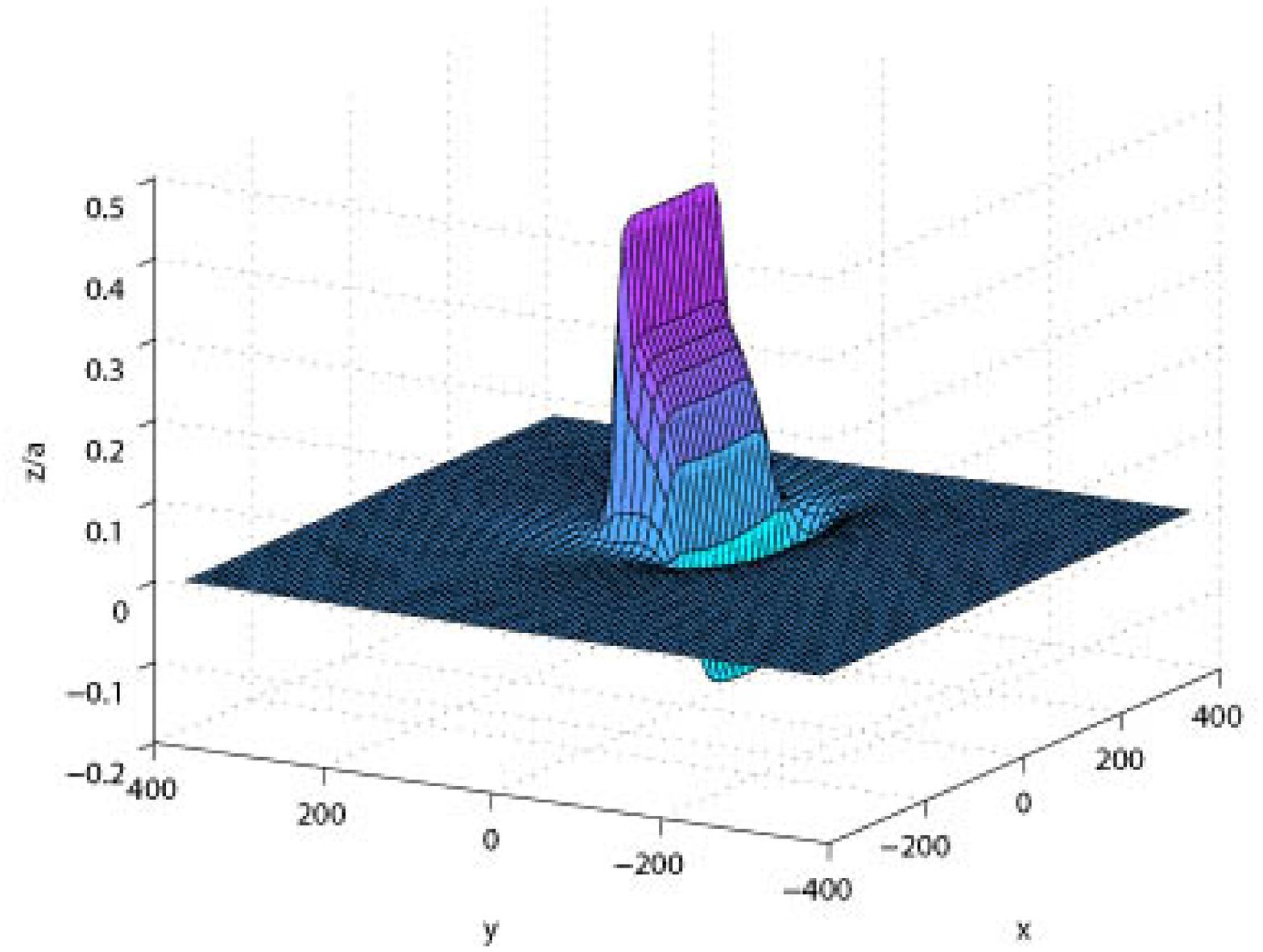}\\
  \caption{Dimensionless free-surface deformation $z/a$ due to dip-slip faulting: $\phi=0$,
$\theta=\pi/2$, $\Dv=(0,U_2,0)$. Here $a$
is $|\Dv|$ (15 m in the present application). The horizontal distances $x$ and $y$ are expressed in kilometers.}
  \label{fig:dip}
\end{figure}

\begin{figure}
  \includegraphics[width=\linewidth]{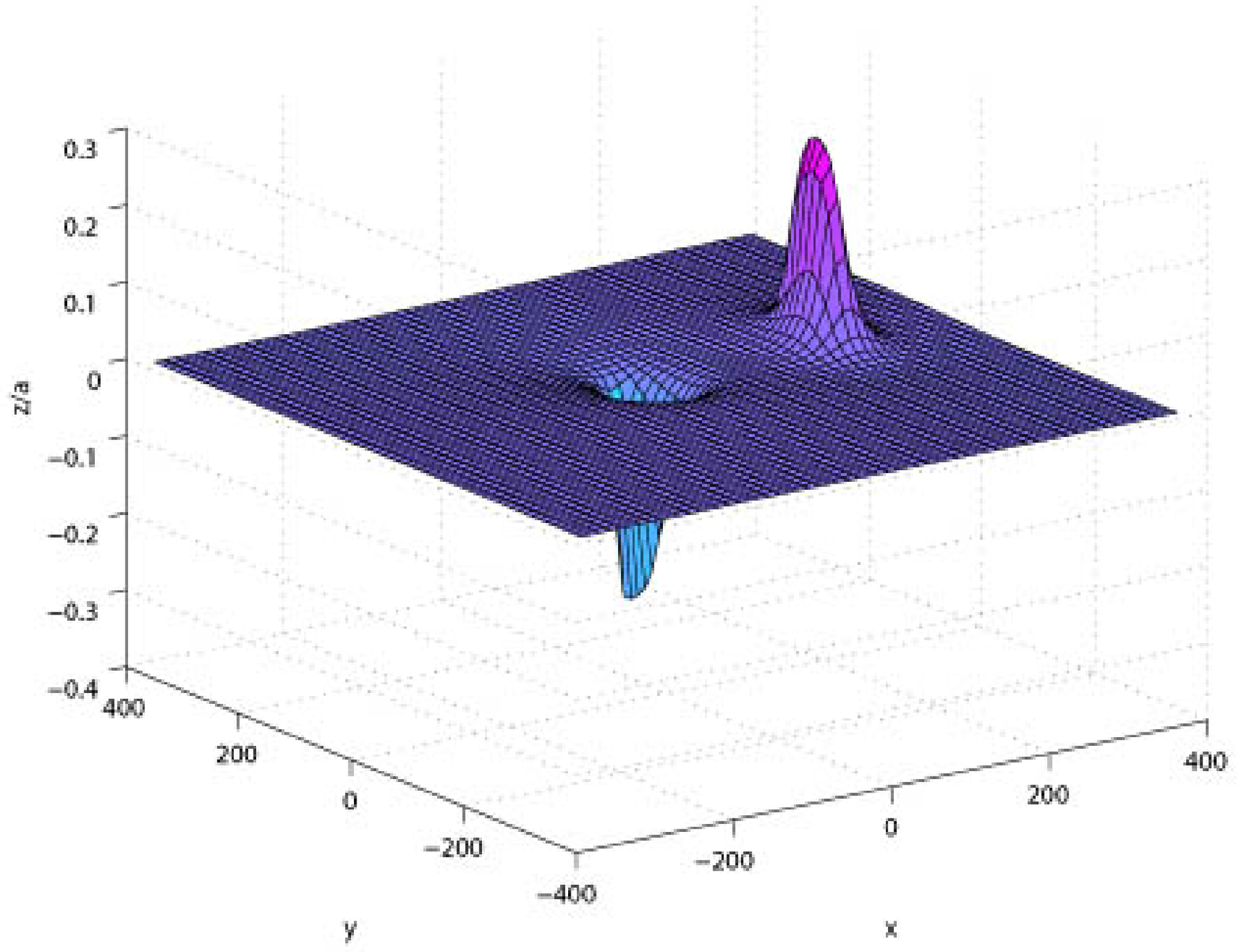}\\
  \caption{Dimensionless free-surface deformation $z/a$ due to strike-slip faulting: $\phi=0$,
$\theta=0$, $\Dv=(U_1,0,0)$. Here $a$
is $|\Dv|$ (15 m in the present application). The horizontal distances $x$ and $y$ are expressed in kilometers.}
  \label{fig:strike}
\end{figure}

\begin{figure}
  \includegraphics[width=0.9\linewidth]{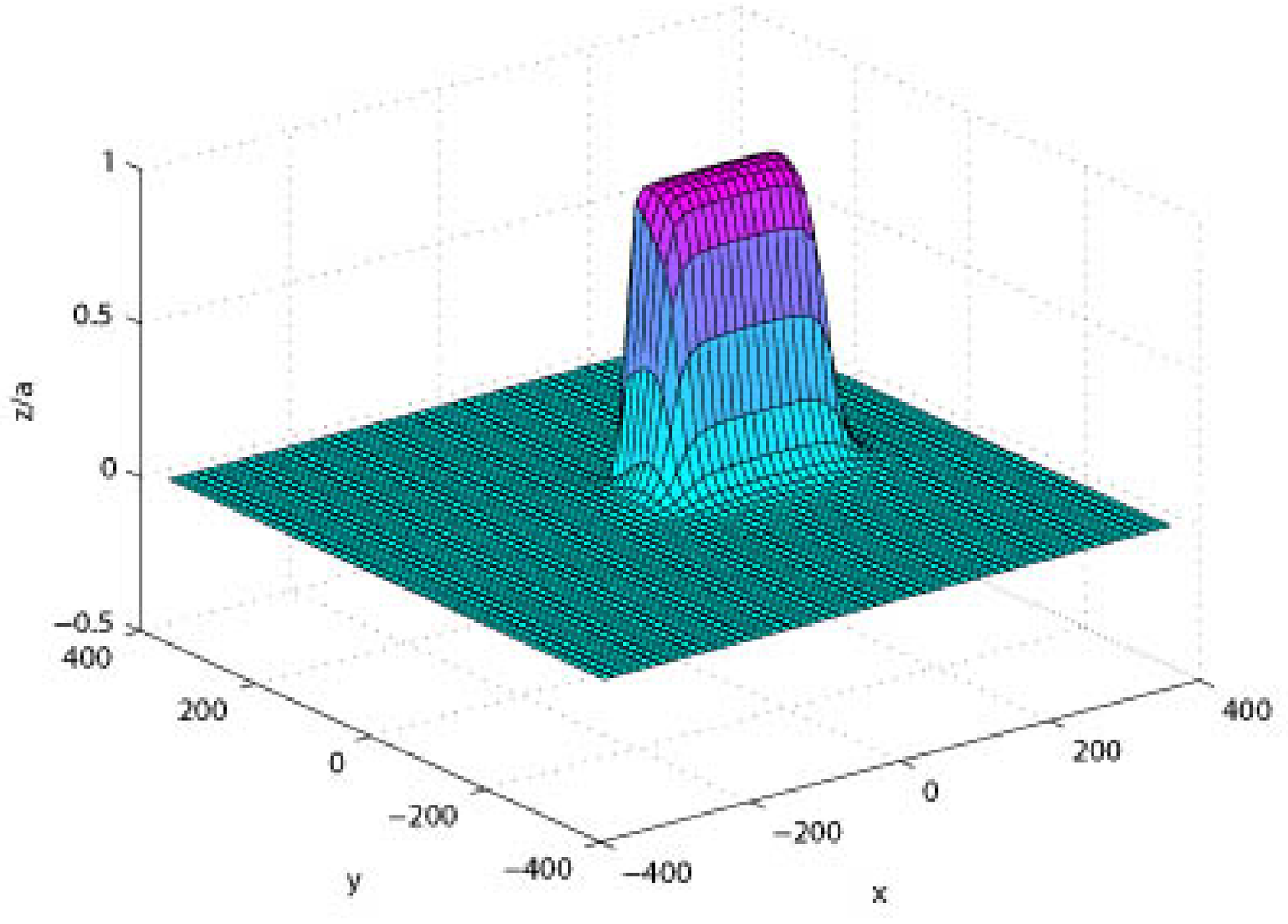}\\
  \caption{Dimensionless free-surface deformation $z/a$ due to tensile faulting: $\phi=\pi/2$,
$\Dv=(0,0,U_3)$. Here $a$ is $|\Dv|$. The horizontal distances $x$ and $y$ are expressed in kilometers.}
  \label{fig:tensile}
\end{figure}

Figures \ref{fig:dip}, \ref{fig:strike}, and \ref{fig:tensile} show
the free-surface deformation due to the three elementary dislocations.
The values of the parameters are given in Table \ref{parset}.
\begin{table}
\centering
\begin{tabular}{lc}
  \hline
{\it parameter} & {\it value} \\
\hline
  Dip angle $\delta$ & $13^\circ$ \\
  Fault depth $d$, km & 25 \\
  Fault length $L$, km & 220 \\
  Fault width $W$, km & 90 \\
  $U_i$, m & 15 \\
  Young modulus $E$, GPa & 9.5 \\
  Poisson's ratio $\nu$ & 0.23 \\
\hline
\end{tabular}
\caption[]{Parameter set used in Figures \ref{fig:dip},
\ref{fig:strike}, and \ref{fig:tensile}.} \label{parset}
\end{table}

\subsection{Curvilinear fault}

In the previous subsection analytical formulas for the free-surface
deformation in the special case of a rectangular fault were given. In
fact, Volterra's formula (\ref{volt2}) allows to
evaluate the displacement field that accompanies fault events with much more
general geometry. The shape of the fault and Burger's vector
are suggested by seismologists and after numerical integration one
can obtain the deformation of the seafloor for more general types of events as well.

\begin{figure}
  \includegraphics[width=0.8\linewidth]{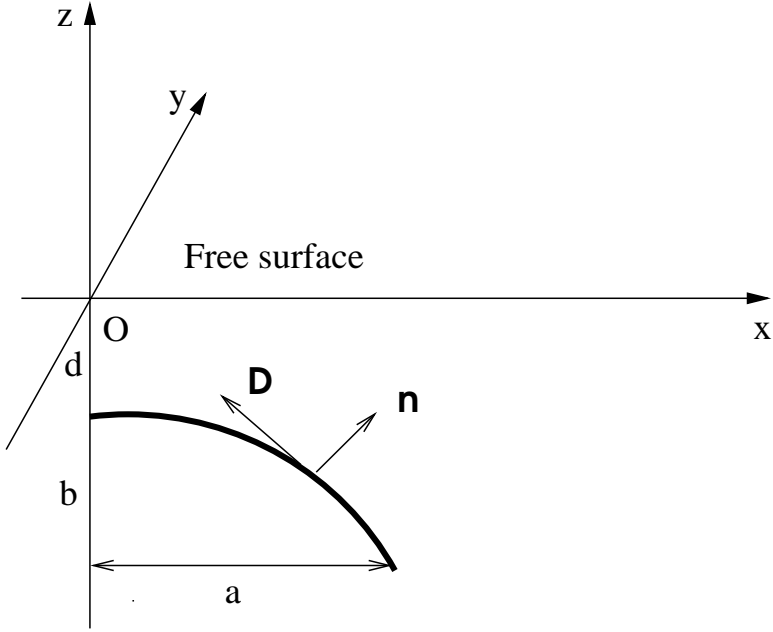}\\
  \caption{Geometry of a fault with elliptical shape.}
  \label{fig:ellipse}
\end{figure}

Here we will consider the case of a fault whose geometry is described by an elliptical
arc (see Figure \ref{fig:ellipse}).
The parametric equations of this surface are given by
$$
  x(\xi,\eta) = \xi, \quad 0\leq\xi\leq a, \qquad y(\xi,\eta) = \eta, \quad -\frac{c}2\leq\eta\leq\frac{c}2, 
$$
$$
  z(\xi,\eta) = -(b+d) + \frac{b}{a}\sqrt{a^2-\xi^2}.
$$
Then the unit normal to this surface can be easily calculated:
$$
    \nv = \left(
    \frac{b\xi}{\sqrt{a^4 + (b^2-a^2)\xi^2}}, 0, \frac{a\sqrt{a^2-\xi^2}}
    {\sqrt{a^4 + (b^2-a^2)\xi^2}}
    \right).
$$
We also need to compute the coefficients of the first fundamental
form in order to reduce the surface integral in (\ref{volt2}) to a
double Riemann integral. These coefficients are
$$ E = \frac{a^4 + \xi^2(b^2-a^2)}{a^2(a^2-\xi^2)}, \quad F = 0, \quad G = 1 $$
and the surface element $dS$ is 
$$ dS = \sqrt{EG-F^2} \,d\xi d\eta = \frac1a\frac{\sqrt{a^4+\xi^2(b^2-a^2)}}{\sqrt{a^2-\xi^2}}
  \, d\xi d\eta. $$

Since in the crust the hydrostatic pressure is very large, it is
natural to impose the condition that $\Dv \cdot \nv = 0.$
The physical meaning of this condition is that both sides of the
fault slide and do not detach. This condition is obviously satisfied
if we take Burger's vector as
$$
    \Dv = D\left(
    \frac{a\sqrt{a^2-\xi^2}}{\sqrt{a^4+\xi^2(b^2-a^2)}}, 0,
    -\frac{b\xi}{\sqrt{a^4+\xi^2(b^2-a^2)}}
    \right).
$$
It is evident that $D = |\Dv|$.

\begin{figure}
  \includegraphics[width=\linewidth]{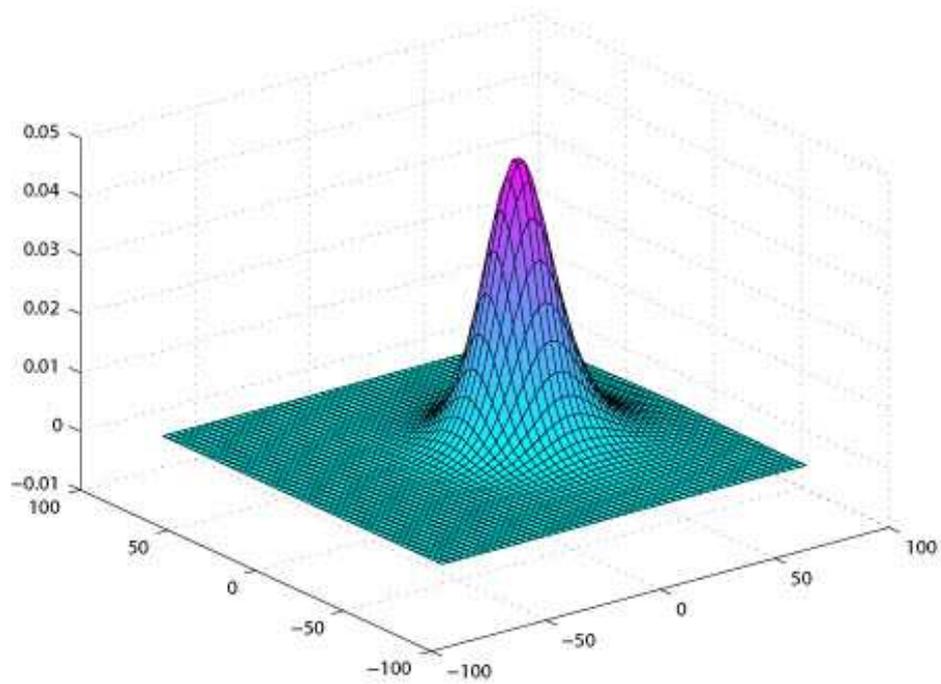}\\
  \caption{Free-surface deformation due to curvilinear faulting. The horizontal distances $x$ and $y$ are expressed in kilometers.}
  \label{fig:ellres}
\end{figure}

The numerical integration was performed using a $9$-point
two-dimensional Gauss-type integration formula.
The result is presented on Figure
\ref{fig:ellres}. The parameter values are given in Table \ref{parellipse}.
\begin{table}
\centering
\begin{tabular}{lc}
  \hline
  {\it parameter} & {\it value} \\
  \hline
  Depth event $d$, km & 20 \\
  Ellipse semiminor axis $a$, km & 17 \\
  Ellipse semimajor axis $b$, km & 6 \\
  Fault width $c$, km & 15 \\
  Young modulus $E$, GPa & 9.5 \\
  Poisson's ratio $\nu$ & 0.23 \\
  \hline
\end{tabular}
\caption[]{Parameter set used in Figure \ref{fig:ellres}.} \label{parellipse}
\end{table}

The example considered in this subsection may not be
physically relevant. However it shows how Okada's solution can be extended. For a more precise modeling of
the faulting event we need to have more information about the earthquake source
and its related parameters. 

After having reviewed the description of the source, we now switch to the deformation of the ocean surface
following a submarine earthquake.
The traditional approach for hydrodynamic modelers is to use
elastic models similar to the model we just described with the seismic
parameters as input in order to evaluate the details of the seafloor deformation.
Then this deformation is translated to the free surface of the ocean and serves as initial condition of the
evolution problem described in the next section.

\section{Solution in fluid domain}

\begin{figure}
\centering
  \includegraphics[width=0.8\linewidth]{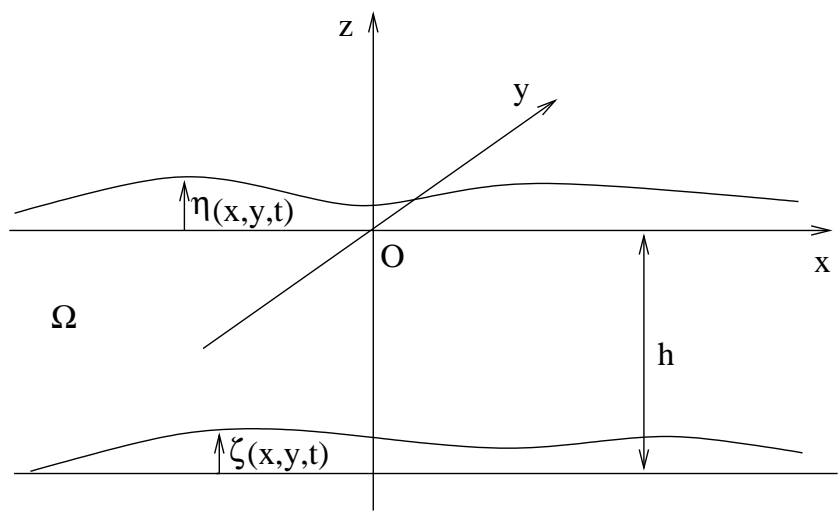}\\
  \caption{Definition of the fluid domain and coordinate system}
  \label{fig:fluid}
\end{figure}

The fluid domain is supposed to represent the ocean above the fault area. 
Let us consider the fluid domain $\Omega$ shown in Figure
\ref{fig:fluid}. It is bounded above by the free surface of the ocean and below
by the rigid ocean floor. The domain $\Omega$ is unbounded in the horizontal
directions $x$ and $y$, and can be written as
$$
  \Omega = \mathbb{R}^2\times\left[-h+\zeta(x,y,t),\eta(x,y,t)\right].
$$
Initially the fluid is assumed to be at rest and the sea bottom to be 
horizontal. Thus, at time $t=0$, the
free surface and the sea bottom are defined by $z=0$ and $z=-h$,
respectively. For time $t>0$ the bottom boundary moves in a
prescribed manner which is given by
$$
    z = -h + \zeta(x,y,t).
$$
The displacement of the sea bottom is assumed to have all the properties
required to compute its Fourier transform in $x,y$ and its Laplace transform in $t$. 
The resulting deformation of the free surface
$z=\eta(x,y,t)$ must be found. It is also assumed that the
fluid is incompressible and the flow is irrotational. The latter
implies the existence of a velocity potential $\phi(x,y,z,t)$ which
completely describes this flow. By definition of $\phi$, the fluid
velocity vector can be expressed as $\qv = \nabla\phi$. Thus, the
continuity equation becomes
\begin{equation}
  \nabla\cdot\qv = \Delta\phi = 0, \quad (x,y,z) \in \Omega.
\label{laplacien}
\end{equation}
The potential $\phi(x,y,z,t)$ must also satisfy the following
kinematic boundary conditions on the free-surface and the solid
boundary, respectively:
\begin{eqnarray}
  \pd{\phi}{z} &=& \pd{\eta}{t} + \pd{\phi}{x}\pd{\eta}{x} +
  \pd{\phi}{y}\pd{\eta}{y}, \qquad z=\eta(x,y,t), \label{kscl} \\
  \pd{\phi}{z} &=& \pd{\zeta}{t} + \pd{\phi}{x}\pd{\zeta}{x} +
  \pd{\phi}{y}\pd{\zeta}{y}, \qquad z=-h + \zeta(x,y,t). \label{kbcl}
\end{eqnarray}
Assuming that viscous effects as well as capillary effects can be neglected, the dynamic condition to be 
satisfied on the free surface reads
\begin{equation}
  \pd{\phi}{t} + \frac12|\nabla\phi|^2 + g\eta = 0, \qquad
  z=\eta(x,y,t). \label{dcl}
\end{equation}
As described above, the initial conditions are given by
\begin{equation}\label{initialcond}
  \eta(x,y,0) = 0 \quad \mbox{and} \quad \zeta(x,y,0) = 0.
\end{equation}

The significance of the various terms in the equations is more transparent when the equations are written in dimensionless
variables. The new independent variables are
$$
    \widetilde{x} = \kappa x, \quad \widetilde{y} = \kappa y, \quad \widetilde{z} = \kappa z, \quad
    \widetilde{t} = \sigma t,
$$
where $\kappa$ is a wavenumber and $\sigma$ is a typical frequency. Note that here the same unit length is used in the
horizontal and vertical directions, as opposed to shallow-water theory.

The new dependent variables are
$$
  \widetilde{\eta} = \frac{\eta}a, \quad \widetilde{\zeta} = \frac{\zeta}{a}, \quad
  \widetilde{\phi} = \frac{\kappa}{a\sigma}\phi,
$$
where $a$ is a characteristic wave amplitude. A dimensionless water depth is also introduced:
$$ \widetilde{h} = \kappa h. $$
In dimensionless form, and after dropping the tildes, equations (\ref{laplacien}--\ref{dcl}) become
$$
  \Delta \phi = 0, \qquad (x,y,z) \in \Omega,
$$
\begin{eqnarray*}
  \pd{\phi}{z} &=& \pd{\eta}{t} + \kappa a \left(
  \pd{\phi}{x}\pd{\eta}{x} + \pd{\phi}{y}\pd{\eta}{y}
  \right), \qquad z = \kappa a \, \eta(x,y,t), \\
  \pd{\phi}{z} &=& \pd{\zeta}{t} + \kappa a \left(
  \pd{\phi}{x}\pd{\zeta}{x} + \pd{\phi}{y}\pd{\zeta}{y}
  \right), \qquad z = -h + \kappa a \, \zeta(x,y,t),
\end{eqnarray*}
$$
  \pd{\phi}{t} + \frac12 \kappa a |\nabla\phi|^2 + \frac{g\kappa}{\sigma^2}
  \eta = 0, \qquad z = \kappa a \, \eta(x,y,t).
$$

Finding the solution to this problem is quite a difficult task due to the
nonlinearities and the a priori unknown free surface. In this study
we linearize the equations and the boundary conditions by taking the limit as $\kappa a\to 0$. 
In fact, the linearized problem can be found
by expanding the unknown functions as power series of a small parameter
$\varepsilon:=\kappa a$. Collecting the lowest order terms in
$\varepsilon$ yields the linear approximation. For the sake of convenience,
we now switch back to the physical variables. The linearized
problem in dimensional variables reads
\begin{equation}\label{lapl}
  \Delta \phi = 0, \qquad (x,y,z) \in \mathbb{R}^2\times[-h, 0],
\end{equation}
\begin{equation}\label{kinfreesurf}
  \pd{\phi}{z} = \pd{\eta}{t}, \qquad z = 0,
\end{equation}
\begin{equation}\label{kinsolb}
  \pd{\phi}{z} = \pd{\zeta}{t}, \qquad z = -h,
\end{equation}
\begin{equation}\label{dynfreesurf}
  \pd{\phi}{t} + g\eta = 0, \qquad z = 0.
\end{equation}

Combining equations (\ref{kinfreesurf}) and (\ref{dynfreesurf})
yields the single free-surface condition
\begin{equation}\label{singlefreesurf}
  \pd{^2\phi}{t^2} + g\pd{\phi}{z} = 0,
  \qquad z = 0.
\end{equation}

This problem will be solved by using the method of integral transforms. We apply
the Fourier transform in $(x,y)$:
\begin{eqnarray*}
  \mathfrak{F}[f] & = & \widehat{f}(k,\ell) = \int\limits_{\mathbb{R}^2} f(x,y)
  e^{-i(kx+\ell y)}\,dx dy, \\
  \mathfrak{F}^{-1}[\widehat{f}] & = & f(x,y) = \frac1{(2\pi)^2}
  \int\limits_{\mathbb{R}^2} \widehat{f}(k,\ell)
  e^{i(kx+\ell y)}\,dk d\ell,
\end{eqnarray*}
and the Laplace transform in time $t$:
$$
  \mathfrak{L}[g] = \tens{g}(s) = \int\limits_0^{+\infty}
  g(t) e^{-st}\, dt.
$$ 
For the combined Fourier and Laplace transforms, the following notation is introduced:
$$ \mathfrak{F}\mathfrak{L}[F(x,y,t)] = \overline{F}(k,\ell,s) = \int\limits_{\mathbb{R}^2}
e^{-i(kx+\ell y)}\,dx dy \int\limits_0^{+\infty} F(x,y,t)
  e^{-st}\, dt. $$ 
After applying the transforms, equations (\ref{lapl}), (\ref{kinsolb}) and
(\ref{singlefreesurf}) become
\begin{equation}\label{lapltrans}
  \od{^2\overline{\phi}}{z^2} - (k^2+\ell^2)\overline{\phi} = 0,
\end{equation}
\begin{equation}\label{kinsolbtrans}
  \od{\overline{\phi}}{z}(k,\ell,-h,s) = s\overline{\zeta}(k,\ell,s),
\end{equation}
\begin{equation}\label{freesurftrans}
  s^2\overline{\phi}(k,\ell,0,s) + g\od{\overline{\phi}}{z} (k,\ell,0,s) = 0.
\end{equation}
The transformed free-surface elevation can be obtained from
(\ref{dynfreesurf}):
\begin{equation}\label{transfreesurf}
  \overline{\eta}(k,\ell,s) = -\frac{s}{g}\overline{\phi}(k,\ell,0,s).
\end{equation}

\begin{figure}
  \includegraphics[width=0.9\linewidth]{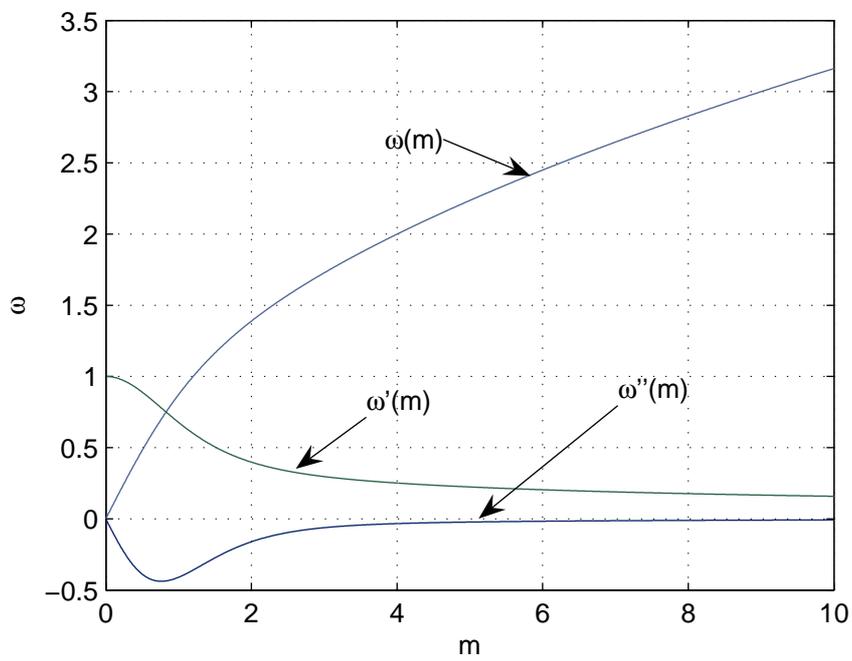}\\
  \caption{Plot of the frequency $\omega(m)=\sqrt{gm\tanh(mh)}$ and its
  derivatives $d\omega/dm$, $d^2\omega/dm^2$. The acceleration due to gravity $g$ and the water depth $h$ have been
set equal to 1.}
  \label{fig:omega}
\end{figure}

A general solution of equation (\ref{lapltrans}) is given by
\begin{equation}\label{genlapltrans}
  \overline{\phi}(k,\ell,z,s) = A(k,\ell,s)\cosh(mz) + B(k,\ell,s)\sinh(mz),
\end{equation}
where $m = \sqrt{k^2+\ell^2}$. The functions $A(k,\ell,s)$ and $B(k,\ell,s)$
can be easily found from the boundary conditions (\ref{kinsolbtrans}) and
(\ref{freesurftrans}):
\begin{eqnarray*}
  A(k,\ell,s) & = & -\frac{gs\overline{\zeta}(k,\ell,s)}{\cosh(mh)[s^2+gm\tanh(mh)]}, \\
  B(k,\ell,s) & = & \frac{s^3\overline{\zeta}(k,\ell,s)}{m\cosh(mh)[s^2+gm\tanh(mh)]}.
\end{eqnarray*}
From now on, the notation
\begin{equation}
\omega = \sqrt{gm\tanh(mh)}
\label{disprelation}
\end{equation}
will be used. The graphs of $\omega(m)$, $\omega'(m)$ and $\omega''(m)$ are shown in Figure \ref{fig:omega}.

Substituting the expressions for the functions $A$, $B$ in (\ref{genlapltrans}) yields
\begin{equation}
  \overline{\phi}(k,\ell,z,s) = -\frac{gs\overline{\zeta}(k,\ell,s)}
  {\cosh(mh)(s^2+\omega^2)}
  \left(\cosh(mz) - \frac{s^2}{gm}\sinh(mz)\right). \label{phihat}
\end{equation}

\subsection{Free-surface elevation}

From (\ref{transfreesurf}), the free-surface elevation becomes
$$
  \overline{\eta}(k,\ell,s) = \frac{s^2\overline{\zeta}(k,\ell,s)}
  {\cosh(mh)(s^2+\omega^2)}.
$$

Inverting the Laplace and Fourier transforms provides the general
integral solution
\begin{equation}\label{genintsol}
  \eta(x,y,t) = \frac{1}{(2\pi)^2}\int\!\!\!\int\limits_{\!\!\!\!\!\R^2}
  \frac{e^{i(kx+\ell y)}}{\cosh(mh)}\frac1{2\pi i}
  \int\limits_{\mu-i\infty}^{\mu+i\infty}
  \frac{s^2\overline{\zeta}(k,\ell,s)}{s^2+\omega^2}e^{st}ds\; dk d\ell.
\end{equation}
One can evaluate the Laplace integral in (\ref{genintsol})
using the convolution theorem:
$$
  \mathfrak{L}[f_1(t)*f_2(t)] = \tens{f_1}(s) \tens{f_2}(s).
$$
It yields
$$
  \eta(x,y,t) = \frac1{(2\pi)^2}\int\!\!\!\int\limits_{\!\!\!\!\!\R^2}
  \frac{e^{i(kx+\ell y)}}{\cosh(mh)}
  \int\limits_0^t (1-\omega\sin\omega\tau)\overline{\zeta}(k,\ell,t-\tau)d\tau \, dk d\ell.
$$

This general solution contains as a special case the
solution for an axisymmetric problem, which we now describe in detail. 
Assume that the initial solid boundary deformation is axisymmetric:
$$
    \zeta(x,y) = \zeta(r), \qquad r = \sqrt{x^2+y^2}.
$$
The Fourier transform
$\mathfrak{F}[\zeta(x,y)] = \widehat{\zeta}(k,\ell)$ of an axisymmetric
function is also axisymmetric with respect to transformation
parameters, i.e.
$$
    \widehat{\zeta}(k,\ell) = \widehat{\zeta}(m), \qquad m := \sqrt{k^2+\ell^2}.
$$
In the following calculation, we use the notation $\psi = \arctan(\ell/k)$. One has
\begin{multline*}
  \widehat{\zeta}(k,\ell) = \int\!\!\!\int\limits_{\!\!\!\!\!\R^2}
  \zeta(r)e^{-i(kx+\ell y)}\;dx dy = \int\limits_0^{2\pi}d\phi
  \int\limits_0^{\infty}\zeta(r)e^{-ir(k\cos\phi+\ell\sin\phi)}rdr =
  \\=\int\limits_0^{2\pi}d\phi\int\limits_0^{\infty} r\zeta(r)
  e^{-irm\cos(\phi-\psi)}dr = \int\limits_0^{\infty} r\zeta(r)
  dr \int\limits_0^\pi (e^{-irm\cos\phi} +
  e^{irm\cos\phi})d\phi. 
\end{multline*}
Using an integral representation of Bessel functions \cite{gradshteyn} finally yields
$$ \widehat{\zeta}(k,\ell) = 2\pi\int\limits_0^{\infty}r\zeta(r)J_0(mr)dr \equiv
  \widehat{\zeta}(m). $$
It follows that
\begin{eqnarray*}
  \eta(r,t) & = & \frac1{(2\pi)^2} \int\limits_0^{2\pi}\;d\psi
  \int\limits_0^{+\infty}\frac{m e^{im r\cos(\phi-\psi)}}
  {\cosh(m h)}dm\int\limits_0^t(1-\omega\sin\omega\tau)
  \overline{\zeta}(m,t-\tau)\;d\tau \\
  & = & \frac1{2\pi}\int\limits_0^{+\infty}
  m\frac{J_0(m r)}{\cosh(m h)}dm
  \int\limits_0^t(1-\omega\sin\omega\tau)\overline{\zeta}(m,t-\tau)d\tau.
\end{eqnarray*}
The last equation gives the general integral solution of the problem
in the case of an axisymmetric seabed deformation. Below we no longer
make this assumption since Okada's solution does not have this
property.

In the present study we consider seabed deformations with
the following structure:
\begin{equation}\label{specrepr}
  \zeta(x,y,t) := \zeta(x,y)T(t).
\end{equation}
Mathematically we separate the time dependence from the spatial coordinates.
There are two main reasons for doing this. First of all we want to
be able to invert analytically the Laplace transform. The
second reason is more fundamental. In fact, dynamic
source models are not easily available. Okada's solution, which was described in the previous section,
provides the static sea-bed deformation $\zeta_0(x,y)$ and we will
consider different time dependencies $T(t)$ to model the time evolution
of the source. Four scenarios will be considered:
\begin{enumerate}
    \item {\bf Instantaneous}: $T_i(t) = H(t)$, where $H(t)$ denotes
    the Heaviside step function,
    \item {\bf Exponential}: $$
    T_e(t) = \left\{%
    \begin{array}{ll}
        0, & t<0, \\
        1 - e^{-\alpha t}, & t\geq 0, 
    \end{array}%
    \right. \quad \mbox{with} \;\; \alpha > 0,
    $$
    \item {\bf Trigonometric}: $T_c(t) = H(t-t_0) + \frac12[1-\cos(\pi t/t_0)]H(t_0-t),$
    \item {\bf Linear}: $$
      T_l(t) = \left\{%
\begin{array}{ll}
    0, & t<0, \\
    t/t_0, & 0\leq t\leq t_0, \\
    1, & t>t_0. \\
\end{array}%
\right.
    $$
\end{enumerate}
\begin{figure}
\centering
  \includegraphics[width=0.8\linewidth]{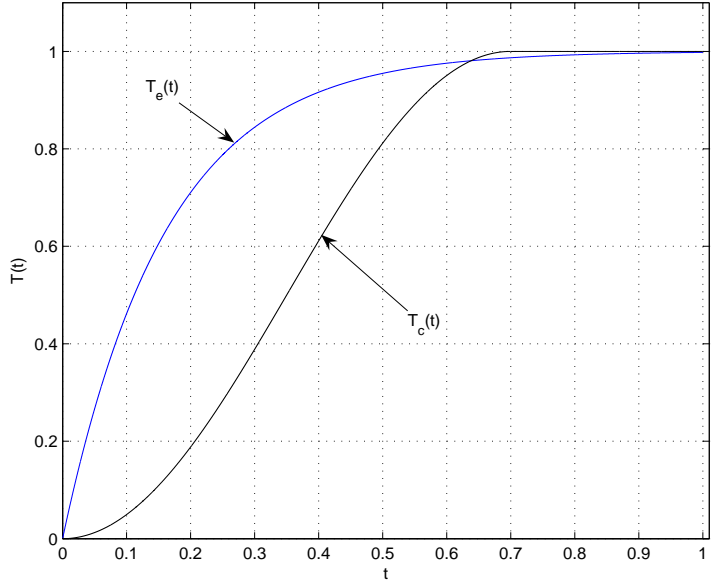}\\
  \caption{Typical graphs of $T_e(t)$ and $T_c(t)$.
  Here we have set $\alpha=6.2$, $t_0=0.7$.}
  \label{fig:time}
\end{figure}
The typical graphs of $T_c(t)$ and $T_e(t)$ are shown in Figure \ref{fig:time}.
Inserting (\ref{specrepr}) into (\ref{genintsol}) yields
\begin{equation}\label{specintsol}
  \eta(x,y,t) = \frac{1}{(2\pi)^2}\int\!\!\!\int\limits_{\!\!\!\!\!\R^2}
  \frac{\widehat{\zeta}(k,\ell)e^{i(kx+\ell y)}}{\cosh(m h)}\frac1{2\pi i}
  \int\limits_{\mu-i\infty}^{\mu+i\infty}
  \frac{s^2\tens{T}(s)}{s^2+\omega^2}e^{st}ds\; dk d\ell.
\end{equation}

Clearly, $\eta(x,y,t)$ depends continuously on the source $\zeta(x,y)$. Physically it means that 
small variations of $\zeta$ (in a reasonable space of functions such as $L^2$) yield small variations of $\eta$. 
Mathematically this problem is said to be well-posed, and this property is essential for modelling the physical 
processes, since it means that small modifications of the ground motion (for example, the error in measurements) 
do not induce huge modifications of the wave patterns.

Using the special representation (\ref{specrepr}) of seabed
deformation and prescribed time-dependencies, one can compute
analytically the Laplace integral in (\ref{specintsol}). To
perform this integration, we first have to compute the Laplace transform
of $T_{i,e,c,l}(t)$. The results are
$$
  \mathfrak{L}[T_i] = \frac1s,\qquad \mathfrak{L}[T_e] =
  \frac{\alpha}{s(\alpha+s)},
$$
$$
  \mathfrak{L}[T_c] =
  (1+e^{-st_0})\frac{\gamma^2}{2s(s^2+\gamma^2)} \;\; \mbox{with} \;\;
  \gamma = \frac{\pi}{t_0}, \qquad
  \mathfrak{L}[T_l] = \frac{1-e^{-st_0}}{t_0s^2}.
$$
Inserting these formulas into the inverse Laplace integral
yields
\begin{eqnarray*}
  \frac1{2\pi i}
  \int\limits_{\mu-i\infty}^{\mu+i\infty}
  \frac{e^{st}s^2\tens{T_i}(s)}{s^2+\omega^2}ds & = &
  \cos\omega t, \\
  \frac1{2\pi i}
  \int\limits_{\mu-i\infty}^{\mu+i\infty}
  \frac{e^{st}s^2\tens{T_e}(s)}{s^2+\omega^2}ds & = &
  -\frac{\alpha^2}{\alpha^2+\omega^2}\left(
  e^{-\alpha t} - \cos\omega t - \frac{\omega}{\alpha}\sin\omega t
  \right), \\
  \frac1{2\pi i}
  \int\limits_{\mu-i\infty}^{\mu+i\infty}
  \frac{e^{st}s^2\tens{T_c}(s)}{s^2+\omega^2}ds & = & \frac{\gamma^2}{2(\gamma^2-\omega^2)} \\
  & & \hspace{-0.5cm} \left(
  \cos\omega t - \cos\gamma t + H(t-t_0)[
  \cos\omega(t-t_0) + \cos\gamma t]\right), \\
  \frac1{2\pi i}
  \int\limits_{\mu-i\infty}^{\mu+i\infty}
  \frac{e^{st}s^2\tens{T_l}(s)}{s^2+\omega^2}ds & = &
  \frac{\sin\omega t - H(t-t_0)\sin\omega(t-t_0)}{\omega t_0}.
\end{eqnarray*}

The final integral formulas for the free-surface elevations with
different time dependencies are as follows:
\begin{eqnarray*}
  \eta_i(x,y,t) & = & \frac{1}{(2\pi)^2}\int\!\!\!\int\limits_{\!\!\!\!\!\R^2}
  \frac{\widehat{\zeta}(k,\ell)e^{i(kx+\ell y)}}{\cosh(m h)}\cos\omega t\; dk
  d\ell, \\
  \eta_e(x,y,t) & = & \frac{-\alpha^2}{(2\pi)^2}\int\!\!\!\int\limits_{\!\!\!\!\!\R^2}
  \frac{\widehat{\zeta}(k,\ell)e^{i(kx+\ell y)}}{\cosh(m h)}
  \left(\frac{e^{-\alpha t} - \cos\omega t - \frac{\omega}{\alpha}\sin\omega t}
  {\alpha^2+\omega^2}\right)\; dk d\ell,
\end{eqnarray*}
\begin{eqnarray*}
  \eta_c(x,y,t) & = & \frac{\gamma^2}{(2\pi)^2}\int\!\!\!\int\limits_{\!\!\!\!\!\R^2}
  \frac{\widehat{\zeta}(k,\ell)e^{i(kx+\ell y)}}{2(\gamma^2-\omega^2)\cosh(m h)}
  \\ & & \hspace{5mm} \left(\cos\omega t - \cos\gamma t+ H(t-t_0)[
  \cos\omega(t-t_0) + \cos\gamma t]\right)\; dk d\ell, \\
\eta_l(x,y,t) & = & \frac{1}{(2\pi)^2}\int\!\!\!\int\limits_{\!\!\!\!\!\R^2}
  \frac{\widehat{\zeta}(k,\ell)e^{i(kx+\ell y)}}{\cosh(m h)}
  \left(\frac{\sin\omega t - H(t-t_0)\sin\omega(t-t_0)}{\omega t_0}\right)\; dk d\ell.
\end{eqnarray*}

\subsection{Velocity field}
In some applications it is important to know not only the
free-surface elevation but also the velocity field in the fluid
domain. One of the goals of this work is to provide an initial
condition for tsunami propagation codes. For the time being,
tsunami modelers take initial seabed deformations and translate them
directly to the free surface in order to obtain the initial condition $\eta(x,y,0)$. Since a priori there is no
information on the flow velocities, they take a zero velocity
field as initial condition for the velocity: $\nabla\phi(x,y,z,0)=0$. The present computations show that it
is indeed a very good approximation if the generation time is short.

In equation (\ref{phihat}), we obtained the Fourier transform of the velocity potential
$\phi(x,y,z,t)$:
\begin{equation}\label{potential}
  \overline{\phi}(k,\ell,z,s) = -\frac{gs\widehat{\zeta}(k,\ell)\tens{T}(s)}
  {\cosh(m h)(s^2+\omega^2)}
  \left(\cosh(m z) - \frac{s^2}{gm}\sinh(m z)\right).
\end{equation}

Let us evaluate the velocity field at an arbitrary level $z=\beta h$
with $-1\leq\beta\leq 0$. In the linear approximation the
value $\beta = 0$ corresponds to the free surface while $\beta=-1$ corresponds to the
bottom. Next we introduce some notation. The horizontal
velocities are denoted by $\uv$. The horizontal gradient $(\partial/\partial x,\partial/\partial y)$ is denoted by
$\nabla_h$. The vertical velocity component is simply $w$. The Fourier transform parameters are denoted 
$\kv = (k,\ell)$.

Taking the Fourier and Laplace transforms of 
$$
  \uv(x,y,t) = \left.\nabla_h \phi(x,y,z,t)\right|_{z=\beta h}
$$
yields
\begin{eqnarray*}
\overline{\uv}(k,\ell,s) & = & -i\overline{\phi}(k,\ell,\beta h, s)\kv \\
 & = & i\frac{gs\widehat{\zeta}(k,\ell)\tens{T}(s)}
  {\cosh(m h)(s^2+\omega^2)}\left(
  \cosh(\beta mh) - \frac{s^2}{gm}\sinh(\beta mh)
  \right)\kv.
\end{eqnarray*}
Inverting the Fourier and Laplace transforms gives the general formula for the
horizontal velocities:
\begin{eqnarray*}
  \uv(x,y,t) & = &
  \frac{ig}{4\pi^2}\int\!\!\!\int\limits_{\!\!\!\!\!\R^2}
  \frac{\kv \widehat{\zeta}(k,\ell)\cosh(m\beta h) e^{i(kx+\ell y)}}{\cosh(m h)}
  \frac{1}{2\pi i}\int\limits_{\mu-i\infty}^{\mu+i\infty}
  \frac{s\tens{T}(s)e^{st}}{s^2+\omega^2}\;ds\; d\kv \\
  & & -\frac{i}{4\pi^2}\int\!\!\!\int\limits_{\!\!\!\!\!\R^2}
  \frac{\kv \widehat{\zeta}(k,\ell)\sinh(m\beta h) e^{i(kx+\ell y)}}{m\cosh(m h)}
  \frac{1}{2\pi i}\int\limits_{\mu-i\infty}^{\mu+i\infty}
  \frac{s^3\tens{T}(s)e^{st}}{s^2+\omega^2}\;ds\; d\kv.
\end{eqnarray*}

After a few computations, one finds the formulas for the time dependencies $T_i$, $T_e$ and $T_l$. For
simplicity we only give the velocities along the free surface ($\beta=0$):
\begin{eqnarray*}
  \uv_i(x,y,t) & = & \frac{ig}{4\pi^2}
 \int\!\!\!\int\limits_{\!\!\!\!\!\R^2}
  \frac{\kv\widehat{\zeta}(k,\ell) e^{i(kx+\ell y)}}{\cosh(m h)}
  \frac{\sin\omega t}{\omega}\; d\kv, \\
\uv_e(x,y,t) & = & \frac{ig\alpha}{4\pi^2}
 \int\!\!\!\int\limits_{\!\!\!\!\!\R^2}
  \frac{\kv\widehat{\zeta}(k,\ell) e^{i(kx+\ell y)}}
  {(\alpha^2+\omega^2)\cosh(m h)}
\left(e^{-\alpha t} - \cos\omega t + \frac{\alpha}{\omega}\sin\omega t \right) \; d\kv, \\
\uv_l(x,y,t) & = & \frac{ig}{4t_0\pi^2}
  \int\!\!\!\int\limits_{\!\!\!\!\!\R^2}
  \frac{\kv\widehat{\zeta}(k,\ell)e^{i(kx+\ell y)}}{\omega^2\cosh(m h)} \\
  & & \hspace{1cm} \left(1-\cos\omega t - H(t-t_0)[1-\cos\omega(t-t_0)]\right)\;
  d\kv.
\end{eqnarray*}

Next we determine the vertical component of the velocity
$w(x,y,z,t)$. It is easy to obtain the Fourier--Laplace transform
$\overline{w}(k,\ell,z,s)$ by differentiating (\ref{potential}):
$$
  \overline{w}(k,\ell,z,s) = \pd{\overline{\phi}}{z} =
  \frac{sg\widehat{\zeta}(k,\ell)\tens{T}(s)}
  {\cosh(m h)(s^2+\omega^2)}\left(
  \frac{s^2}{g}\cosh(m z) - m\sinh(m z)
  \right).
$$
Inverting this transform yields
\begin{eqnarray*}
  w(x,y,z,t) & = & \frac1{4\pi^2}
  \int\!\!\!\int\limits_{\!\!\!\!\!\R^2}
  \frac{\cosh(m z)\widehat{\zeta}(k,\ell)}{\cosh(m h)}e^{i(kx+\ell y)}
  \frac1{2\pi i}\int\limits_{\mu-i\infty}^{\mu+i\infty}
  \frac{s^3 \tens{T}(s)e^{st}}{s^2+\omega^2}\;ds\;d\kv \\
  & & -\frac{g}{4\pi^2}
  \int\!\!\!\int\limits_{\!\!\!\!\!\R^2}
  \frac{m\sinh(m z)\widehat{\zeta}(k,\ell)}{\cosh(m h)}e^{i(kx+\ell y)}
  \frac1{2\pi i}\int\limits_{\mu-i\infty}^{\mu+i\infty}
  \frac{s\tens{T}(s)e^{st}}{s^2+\omega^2}\;ds\;d\kv,
\end{eqnarray*}
for $-h <z\leq 0$. One can easily obtain the expression of the vertical velocity at a given vertical
level by substituting $z=\beta h$ in the expression for $w$.

The easiest way to compute the vertical velocity $w$ along the free surface
is to use the boundary condition (\ref{kinfreesurf}). Indeed,
the expression for $w$ can be simply derived by differentiating the known formula for $\eta_{i,e,c,l}(x,y,t)$. 
Note that formally the derivative gives the
distributions $\delta(t)$ and $\delta(t-t_0)$ under the integral
sign. It is a consequence of the idealized time
behaviour (such as the instantaneous scenario) and 
it is a disadvantage of the Laplace transform method. In order to avoid
these distributions we can consider the solutions only
for $t>0$ and $t\neq t_0$. From a practical point of view there is no
restriction since for any $\varepsilon > 0$ we can set
$t=\varepsilon$ or $t = t_0 + \varepsilon$. For small values of
$\varepsilon$ this will give a very good approximation of the solution
behaviour at these ``critical'' instants of time. Under this
assumption we give the distribution-free expressions for the vertical velocity along
the free surface:
\begin{eqnarray*}
  w_i(x,y,t) & = & -\frac{1}{4\pi^2}
  \int\!\!\!\int\limits_{\!\!\!\!\!\R^2}
  \frac{\widehat{\zeta}(k,\ell)e^{i(kx+\ell y)}}
  {\cosh(m h)}\omega\sin\omega t
  \; d\kv, \\
  w_e(x,y,t) & = & \frac{\alpha^3}{4\pi^2}
  \int\!\!\!\int\limits_{\!\!\!\!\!\R^2}
  \frac{\widehat{\zeta}(k,\ell)e^{i(kx+\ell y)}}
  {(\alpha^2+\omega^2)\cosh(m h)}
  \left(
  e^{-\alpha t} + \frac{\omega^2}{\alpha^2}\cos\omega t
  - \frac{\omega}{\alpha}\sin\omega t\right)
  \; d\kv, \\
  w_c(x,y,t) & = & -\frac{\gamma^2}{4\pi^2}
  \int\!\!\!\int\limits_{\!\!\!\!\!\R^2}
  \frac{\widehat{\zeta}(k,\ell)e^{i(kx+\ell y)}}
  {2(\gamma^2-\omega^2)\cosh(m h)}
  \bigl(
  \omega\sin\omega t - \gamma\sin\gamma t \\
& & \hspace{2cm} +H(t-t_0)[\omega\sin\omega(t-t_0) + \gamma\sin\gamma t]
  \bigr)\; d\kv, \\
  w_l(x,y,t) & = & \frac1{4t_0\pi^2}
  \int\!\!\!\int\limits_{\!\!\!\!\!\R^2}
  \frac{\widehat{\zeta}(k,\ell)e^{i(kx+\ell y)}}{\cosh(m h)}
  \left[\cos\omega t - H(t-t_0)\cos\omega(t-t_0)\right]\;d\kv.
\end{eqnarray*}

\subsection{Pressure on the bottom}

Since tsunameters have one component that measures the pressure at the bottom (bottom pressure
recorder or simply BPR \cite{Gonz}), it is interesting to 
provide as well the expression $p_b(x,y,t)$ for the pressure at the bottom. The pressure $p(x,y,z,t)$ can be obtained
from Bernoulli's equation, which was written explicitly for the free surface in 
equation (\ref{dcl}), but is valid everywhere in the fluid:
\begin{equation}
  \pd{\phi}{t} + \frac12|\nabla\phi|^2 + gz + \frac{p}{\rho} = 0. 
\label{bernou}
\end{equation}
After linearization, equation (\ref{bernou}) becomes
\begin{equation}
  \pd{\phi}{t} + gz + \frac{p}{\rho} = 0. 
\label{bernou_l}
\end{equation}
Along the bottom, it reduces to
\begin{equation}
  \pd{\phi}{t} + g(-h+\zeta) + \frac{p_b}{\rho} = 0, \qquad z=-h. 
\label{bernou_l_b}
\end{equation}
The time-derivative of the velocity potential is readily available in Fourier space. 
Inverting the Fourier and Laplace transforms and evaluating the resulting expression 
at $z=-h$ gives for the four time scenarios, respectively,
\begin{eqnarray*}
 \pd{\phi_i}{t} & = & -\frac{g}{(2\pi)^2}\int\!\!\!\int\limits_{\!\!\!\!\!\R^2}
 \frac{\widehat{\zeta}(k,\ell)e^{i(kx+\ell y)}}{\cosh^2(mh)}\cos \omega t \; d\kv, \\ 
 \pd{\phi_e}{t} & = & \frac{g\alpha^2}{(2\pi)^2}\int\!\!\!\int\limits_{\!\!\!\!\!\R^2}
 \frac{\widehat{\zeta}(k,\ell)e^{i(kx+\ell y)}}{\alpha^2+\omega^2}
 \left(e^{-\alpha t} - \cos\omega t - \frac{\omega}{\alpha}\sin\omega t\right) \; d\kv + \frac{\alpha^4}{(2\pi)^2} \\ 
 & & \hspace{-0.5cm} \int\!\!\!\int\limits_{\!\!\!\!\!\R^2}
 \frac{\widehat{\zeta}(k,\ell)\tanh(mh)e^{i(kx+\ell y)}}{m(\alpha^2+\omega^2)}
 \left(e^{-\alpha t} + \Bigl(\frac{\omega}{\alpha}\Bigr)^2\cos\omega t + 
 \Bigl(\frac{\omega}{\alpha}\Bigr)^3\sin\omega t\right)\; d\kv,
\end{eqnarray*}
\begin{eqnarray*}
\pd{\phi_l}{t} & = & -\frac{g}{t_0(2\pi)^2}\int\!\!\!\int\limits_{\!\!\!\!\!\R^2}
 \frac{\widehat{\zeta}(k,\ell)e^{i(kx+\ell y)}}{\omega\cosh^2(mh)}
 \left[\sin\omega t - H(t-t_0)\sin\omega(t-t_0)\right] \; d\kv.
\end{eqnarray*}

The bottom pressure deviation from the hydrostatic pressure is then given by
$$ p_b(x,y,t) = -\left.\rho\pd{\phi}{t}\right|_{z=-h} - \rho g\zeta. $$
Plots of the bottom pressure will be given in Section 4.
 
\subsection{Asymptotic analysis of integral solutions}

In this subsection, we apply the method of stationary
phase in order to estimate the far-field behaviour of the solutions. There is a lot of literature on 
this topic (see for example \cite{erdelyi,murray,petr,BleisHandel,egorov}). This
method is a classical method in asymptotic analysis. To
our knowledge, the stationary phase method was first used by
Kelvin \cite{kelvin} in the context of linear water-wave theory.

The motivation to obtain asymptotic formulas for integral
solutions was mainly due to numerical difficulties to calculate the
solutions for large values of $x$ and $y$. From equation (\ref{genintsol}), it is
clear that the integrand is highly oscillatory. In order to be
able to resolve these oscillations, several
discretization points are needed per period. This becomes extremely expensive
as $r=\sqrt{x^2+y^2}\to \infty$. The numerical method used in the
present study is based on a Filon-type quadrature formula
\cite{filon} and has been adapted to double integrals with $\exp[i(kx+\ell y)]$
oscillations. The idea of this method consists in
interpolating only the amplitude of the integrand at discretization points by
some kind of polynomial or spline and then performing exact
integration for the oscillating part of the integrand. This method seems
to be quite efficient.

Let us first obtain an asymptotic representation for integral
solutions of the general form
\begin{equation}\label{trhomega}
  \eta(x,y,t) = \frac1{4\pi^2}
  \int\!\!\!\int\limits_{\!\!\!\!\!\R^2}
  \frac{\widehat{\zeta}(k,\ell)e^{i(kx+\ell y)}}{\cosh(m h)} T(m,t) \;
  dk d\ell, \;\ m = \sqrt{k^2+\ell^2}.
\end{equation}
Comparing with equation (\ref{specintsol}) shows that $T(m,t)$ is in fact
$$
  T(m,t) = \frac1{2\pi i}
  \int\limits_{\mu-i\infty}^{\mu+i\infty}
  \frac{s^2\tens{T}(s)}{s^2+\omega^2}e^{st}\;ds.
$$
For example, we showed above that for an instantaneous seabed
deformation $T(m,t) = \cos\omega t$, where $\omega^2 =
gm\tanh m h$. For the time being, we do not specify the
time behaviour $\tens{T}(s)$.

In equation (\ref{trhomega}), we switch to polar coordinates $m$ and
$\psi=\arctan(\ell/k)$:
\begin{eqnarray*}
  \eta(x,y,t) & = & \frac1{4\pi^2}
  \int\limits_0^{\infty}
  \int\limits_0^{2\pi}
  \frac{\widehat{\zeta}(m,\psi)e^{im r\cos(\varphi-\psi)}}{\cosh(m h)} T(m,t)m \;
  d\psi dm \\ & = &
  \frac1{4\pi^2}
  \int\limits_0^{\infty}
  \frac{m T(m,t)}{\cosh(m h)}\;dm
  \int\limits_0^{2\pi}
  \widehat{\zeta}(m,\psi)e^{im r\cos(\varphi-\psi)}\;d\psi,
\end{eqnarray*}
where $(r,\varphi)$ are the polar coordinates of $(x,y)$. In the last
expression, the phase function is $\Phi = m r\cos(\varphi-\psi)$.
Stationary phase points satisfy the condition $\partial\Phi/\partial\psi = 0$,
which yields two phases: $\psi_1 = \varphi$ and $\psi_2 = \varphi+\pi$.
An approximation to equation (\ref{trhomega}) is then obtained by applying the
method of stationary phase to the integral over $\psi$:
$$
  \eta(r,\phi,t) \simeq 
  \frac1{\sqrt{8\pi^3r}}\int\limits_0^{\infty}
  \frac{\sqrt{m}T(m,t)}{\cosh(m h)}\left(
  \widehat{\zeta}(m,\varphi)e^{i(\frac{\pi}4-mr)} +
  \widehat{\zeta}(m,\varphi+\pi)e^{i(m r - \frac{\pi}4)}\right)
  dm.
$$
This expression cannot be simplified if we do not make any further hypotheses on the function $T(m,t)$.

Since we are looking for the far field solution
behaviour, the details of wave formation are not important. Thus we
will assume that the initial seabed deformation is instantaneous:
$$
  T(m,t) = \cos\omega t = \frac{e^{i\omega t} + e^{-i\omega t}}{2}.
$$
Inserting this particular function $T(m,t)$ in equation (\ref{trhomega}) yields
$$
  \eta(r,\varphi,t) = \frac{1}{8\pi^2}\bigl(I_1 + I_2\bigr),
$$
where
\begin{eqnarray*}
  I_1 & = & \int\limits_0^{\infty}
  \frac{m\widehat{\zeta}(m,\psi)}{\cosh(m h)}
  \int\limits_0^{2\pi}
  e^{i(\omega t + m r\cos(\varphi - \psi))}\;d\psi dm, \\
  I_2 & = & \int\limits_0^{\infty}
  \frac{m\widehat{\zeta}(m,\psi)}{\cosh(m h)}
  \int\limits_0^{2\pi}
  e^{i(-\omega t + m r\cos(\varphi - \psi))}\;d\psi dm.
\end{eqnarray*}
The stationary phase function in these integrals is
$$
  \Phi (m,\psi) = m r\cos(\varphi - \psi) \pm \omega t,
  \qquad \omega^2(m) = gm\tanh mh.
$$
The points of stationary phase are then obtained from the conditions
$$
  \pd{\Phi}{\psi} = 0, \quad \pd{\Phi}{m} = 0.
$$
The first equation gives two points, $\psi_1 = \varphi$ and $\psi_2 =
\varphi + \pi$, as before. The second condition yields
\begin{equation}\label{definerho}
  \frac{r}{t}\cos(\varphi-\psi_{1,2}) = \mp\od{\omega}{m}.
\end{equation}
Since $d\omega/dm$ decreases from $\sqrt{gh}$ to 0 as
$m$ goes from 0 to $\infty$ (see Figure \ref{fig:omega}), this equation has a unique solution for $m$
if $\abs{r/t} \leq \sqrt{gh}$. This unique solution will be denoted by $m^*$. 

For $\abs{r}> t\sqrt{gh}$, there is no stationary phase. It means physically that the wave has not yet reached this region.
So we can approximately set $I_1 \approx 0$ and $I_2 \approx 0$. From the positivity of the function $d\omega/dm$ one can deduce 
that $\psi_1 = \varphi$ is a stationary phase point only for the integral $I_2$. Similarly, $\psi_2 = \varphi + \pi$ 
is a stationary point only for the integral $I_1$.

Let us obtain an asymptotic formula for the first integral:
\begin{eqnarray*}
  I_1 & \approx & \int\limits_0^{\infty} \frac{m}{\cosh(m h)}
  \left(
  \sqrt{\frac{2\pi}{m r}} \widehat\zeta(m,\varphi+\pi)
  e^{i(\omega t - m r)}e^{i\frac{\pi}4}
  \right)\; dm \\
  & = & \sqrt{\frac{2\pi}{r}}e^{i\frac{\pi}4}
  \int\limits_0^{\infty}
  \frac{\widehat\zeta(m,\varphi+\pi)}{\cosh(m h)}
  \sqrt{m} e^{i(\omega t -m r)} \; dm \\
  & \approx &
  \sqrt{\frac{2\pi}{r}}e^{i\frac{\pi}4}\left(
  \sqrt{\frac{2\pi m^*}{\abs{\omega''(m^*)}t}}
  \frac{\widehat\zeta(m^*,\varphi+\pi)}{\cosh(m^*h)}
  e^{i(\omega(m^*)t - m^* r)}e^{-i\frac{\pi}4}\right) \\
  & = & \frac{2\pi}{t}
  \sqrt{\frac{m^*}{-\omega'' \omega'}}\,
  \frac{\widehat\zeta(m^*,\varphi+\pi)}{\cosh(m^*h)}
  e^{i(\omega(m^*)t - m^* r)}.
\end{eqnarray*}
In this estimate we have used 
equation (\ref{definerho}) evaluated at the stationary phase point $(m^*,\psi_2)$: 
\begin{equation}\label{relrt}
  r = t\left.\od{\omega}{m}\right|_{m=m^*}.
\end{equation}

Similarly one can obtain an estimate for the integral $I_2$:
$$
  I_2 \approx \frac{2\pi}{t}
  \sqrt{\frac{m^*}{-\omega'' \omega'}}\,
  \frac{\widehat\zeta(m^*,\varphi)}{\cosh(m^*h)}
  e^{-i(\omega(m^*)t - m^* r)}.
$$
Asymptotic values have been obtained for the integrals. As is easily observed from the expressions for $I_1$ and $I_2$,
the wave train decays as $1/t$, or $1/r$, which is equivalent since $r$ and $t$ are connected by relation (\ref{relrt}).

\section{Numerical results}
A lot of numerical computations based on the analytical formulas obtained in the previous sections have been performed. 
Because of the lack of information about the real dynamical characteristics of tsunami sources, we cannot really conclude 
which time dependence gives the best description of tsunami generation. At this stage it is still very difficult or even 
impossible. 

Numerical experiments showed that the largest wave amplitudes with the time dependence $T_c(t)$ were obtained for
relatively small values of the characteristic time $t_0$. The exponential dependence has shown higher amplitudes 
for relatively longer characteristic times.
The instantaneous scenario $T_i$ gives at the free surface the initial seabed deformation with a 
slightly lower amplitude (the factor that we obtained was typically about $0.8\sim 0.94$). The water has a high-pass filter effect
on the initial solid boundary deformation. The linear time dependence $T_l(t)$ showed a linear growth of wave amplitude from 
0 to also $\approx 0.9\zeta_0$, where $\zeta_0 = \max\limits_{(x,y)\in\R^2}\abs{ \zeta(x,y)}$. 

\begin{figure}
  \includegraphics[width=1.05\linewidth]{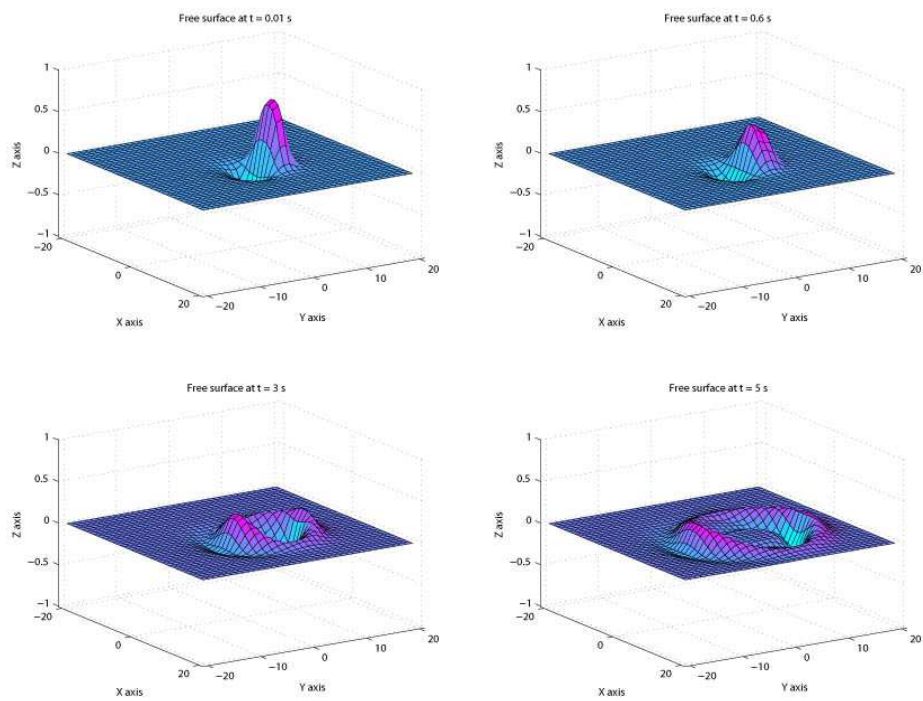}
  \caption{Free-surface elevation at $t=0.01, 0.6, 3, 5$ in dimensionless time. In physical time it corresponds to one second,
one minute, five minutes and eight minutes and a half after the initial seabed deformation.}
  \label{fig:eta0}
\end{figure}

\begin{figure}
  \includegraphics[width=1.05\linewidth]{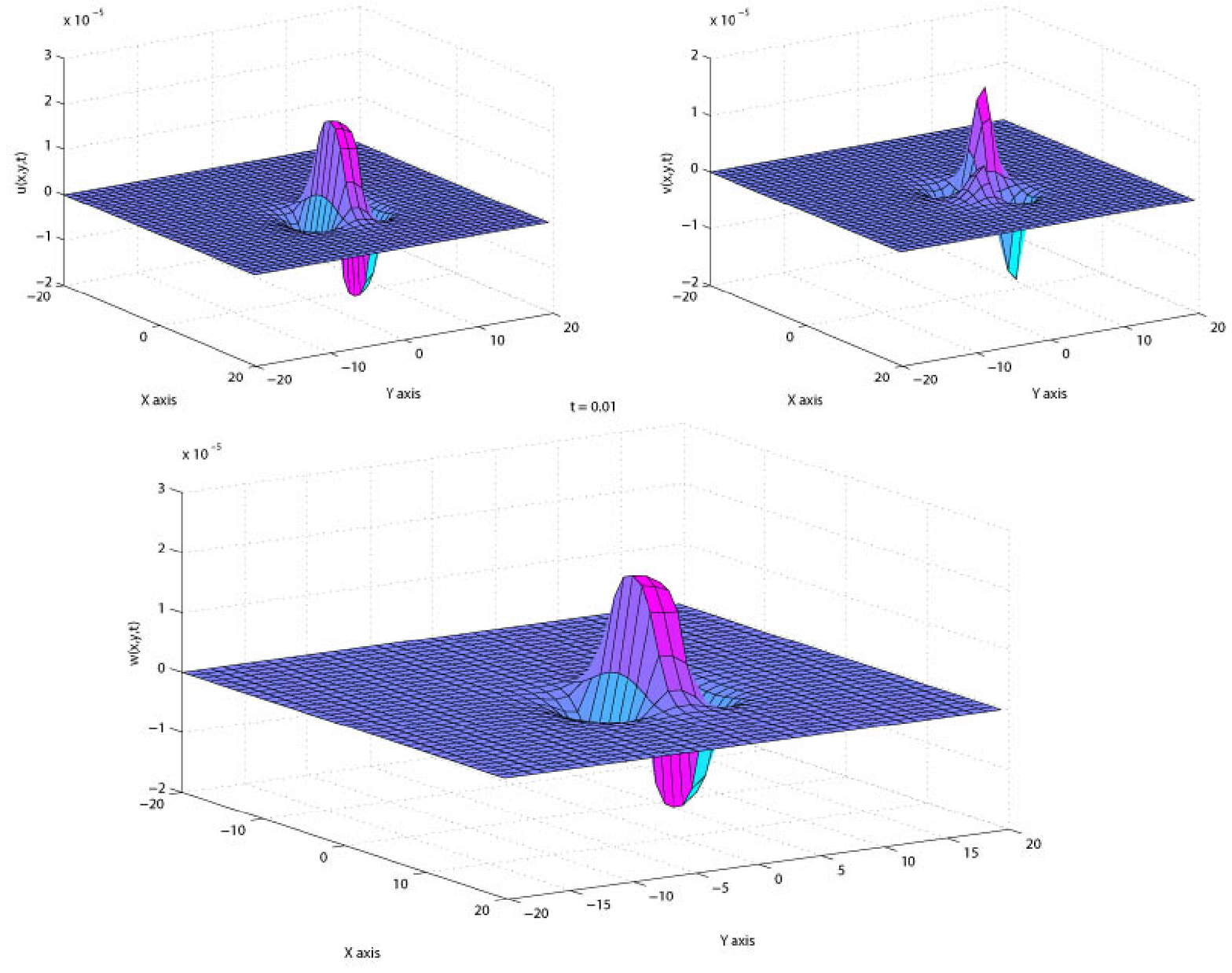}
  \caption{Components $u$, $v$ and $w$ of the velocity field computed along the free surface at $t=0.01$, that is one
second after the initial seabed deformation.}
  \label{fig:uvw}
\end{figure}

\begin{figure}
  \includegraphics[width=1.05\linewidth]{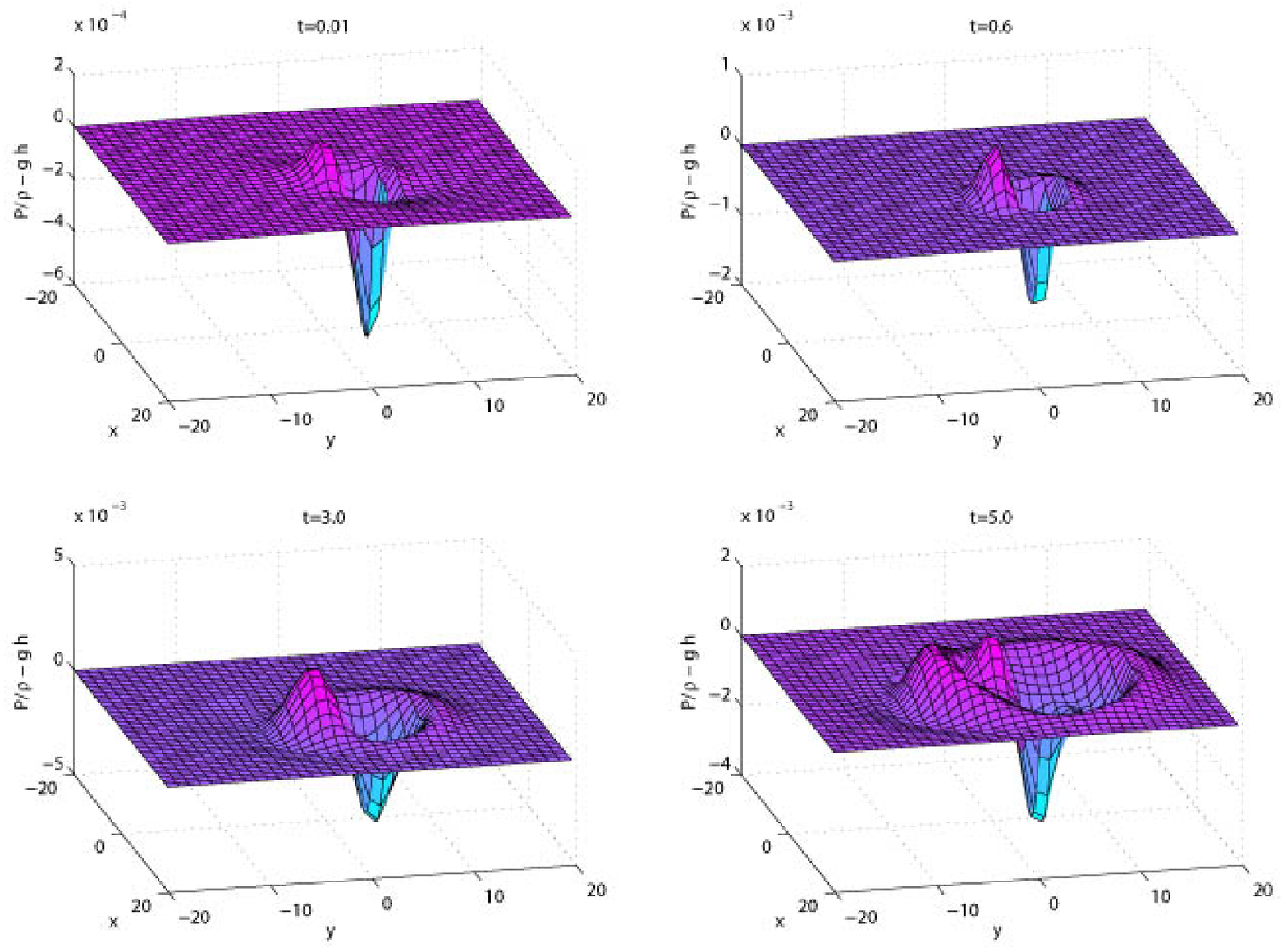}
  \caption{Bottom pressure at $t=0.01, 0.6, 3, 5$ in dimensionless time. In physical time it corresponds to one second,
one minute, five minutes and eight minutes and a half after the initial seabed deformation.}
  \label{fig:press}
\end{figure}

In this section we provide several plots (Figure \ref{fig:eta0}) of the free-surface deformation. 
For illustration purposes, we have chosen the instantaneous seabed deformation since it is the most widely used. 
The values of the parameters used in the computations are given in Table \ref{parameters}. We also give plots 
 of the velocity components on the free surface a few seconds (physical) after the instantaneous deformation
(Figure \ref{fig:uvw}). Finally, plots of the bottom dynamic pressure are given in Figure \ref{fig:press}.

\begin{table}
\begin{center}
\begin{tabular}{lc}
\hline
\bf{Parameter} & \bf{Value} \\
\hline
Young modulus, $E$, GPa & 9.5 \\
Poisson ratio, $\nu$ & 0.27 \\
Fault depth, $d$, km & 20 \\
Dip angle, $\delta$, $^\circ$ & 13 \\
Strike angle, $\theta$, $^\circ$ & 90 \\
Normal angle, $\phi$, $^\circ$ & 0 \\
Fault length, $L$, km & 60 \\
Fault width, $W$, km & 40 \\
Burger's vector length, $\abs{\Dv}$, m & 15 \\
Water depth, $h$, km & 4 \\
Acceleration due to gravity, $g$, $m/s^2$ & 9.8 \\
Wave number, $k$, $1/m$ & $10^{-4}$ \\
Angular frequency, $\omega$, Hz & $10^{-2}$ \\
\hline
\end{tabular}
\caption{Physical parameters used in the numerical computations}
\label{parameters}
\end{center}
\end{table}

From Figure \ref{fig:uvw} it is clear that the velocity field is really negligible 
in the beginning of wave formation. Numerical computations showed that this situation does not change 
if one takes other time-dependences.

The main focus of the present paper is the generation of waves by a moving bottom. 
The asymptotic behaviour of various sets of initial data propagating in a fluid of uniform depth has been studied
in detail by Hammack and Segur \cite{Segur2,Segur3}. In particular, they showed that the behaviours for an 
initial elevation wave and for an initial depression wave are different.


\end{document}